\documentclass[modern,twocolumn]{aastex62}

\graphicspath{{./}{figures/}}

\received{}
\revised{}
\accepted{}
\submitjournal{ApJ}

\shorttitle{}
\shortauthors{}
\begin{document}

\title{A study of sunspot 3 minute oscillations using ALMA and GST}

\correspondingauthor{Yi Chai}
\email{yc389@njit.edu}

\author{Yi Chai}
\affiliation{Center for Solar-Terrestrial Research \\
New Jersey Institute of Technology \\
Newark, NJ 07102, USA}

\author{Dale E. Gary}
\affiliation{Center for Solar-Terrestrial Research \\
New Jersey Institute of Technology \\
Newark, NJ 07102, USA}

\author{Kevin P. Reardon}
\affiliation{National Solar Observatory, Boulder, CO, 80303}
\affiliation{Department of Astrophysical and Planetary Sciences, University of Colorado, Boulder, CO, 80303}

\author{Vasyl Yurchyshyn}
\affiliation{Big Bear Solar Observatory, Center for Solar-Terrestrial Research \\
New Jersey Institute of Technology \\
Newark, NJ 07102, USA}

%% Note that the \and command from previous versions of AASTeX is now
%% depreciated in this version as it is no longer necessary. AASTeX 
%% automatically takes care of all commas and "and"s between authors names.

%% AASTeX 6.2 has the new \collaboration and \nocollaboration commands to
%% provide the collaboration status of a group of authors. These commands 
%% can be used either before or after the list of corresponding authors. The
%% argument for \collaboration is the collaboration identifier. Authors are
%% encouraged to surround collaboration identifiers with ()s. The 
%% \nocollaboration command takes no argument and exists to indicate that
%% the nearby authors are not part of surrounding collaborations.

%% Mark off the abstract in the ``abstract'' environment. 
\begin{abstract}
Waves and oscillations are important solar phenomena, not only because they can propagate and dissipate energy in the chromosphere, but also because they carry information about the structure of the atmosphere in which they propagate. The nature of the three-minute oscillations observed in the umbral region of sunspots is considered to be an effect of propagation of magnetohydrodynamic (MHD) waves upward from below the photosphere. We present a study of sunspot oscillations and wave propagation in NOAA AR 12470 using an approximately one-hour long data set acquired on 2015 December 17 by the Atacama Large Millimeter/submillimeter Array (ALMA), the Goode Solar Telescope (GST) operating at the Big Bear Solar Observatory (BBSO), the Atmospheric Imaging Assembly (AIA) on board the Solar Dynamics Observatory (SDO), and the Interface Region Imaging Spectrograph (IRIS). The ALMA data are unique in providing a time-series of direct temperature measurements in the sunspot chromosphere. The two-second cadence of ALMA images allows us to well resolve the three-minute periods typical of sunspot oscillations in the chromosphere. Fourier analysis is applied to ALMA Band 3 ($\sim$100 GHz, $\sim$3 mm) and GST H$\alpha$ data sets to obtain power spectra as well as oscillation phase information. We analysed properties of the wave propagation by combining multiple wavelengths that probe physical parameters of solar atmosphere at different heights. We find that the ALMA temperature fluctuations are consistent with that expected for a propagating acoustic wave, with a slight asymmetry indicating non-linear steepening.

\end{abstract}

%% Keywords should appear after the \end{abstract} command. 
%% See the online documentation for the full list of available subject
%% keywords and the rules for their use.

%% From the front matter, we move on to the body of the paper.
%% Sections are demarcated by \section and \subsection, respectively.
%% Observe the use of the LaTeX \label
%% command after the \subsection to give a symbolic KEY to the
%% subsection for cross-referencing in a \ref command.
%% You can use LaTeX's \ref and \label commands to keep track of
%% cross-references to sections, equations, tables, and figures.
%% That way, if you change the order of any elements, LaTeX will
%% automatically renumber them.
%%
%% We recommend that authors also use the natbib \citep
%% and \citet commands to identify citations.  The citations are
%% tied to the reference list via symbolic KEYs. The KEY corresponds
%% to the KEY in the \bibitem in the reference list below. 

\section{Introduction}

Sunspot oscillations are a frequently studied wave phenomenon in the solar atmosphere. These oscillations are directly connected with the propagation of solar magnetohydrodynamic (MHD) waves and may play a role in coronal and chromospheric heating. Moreover, they can also serve as a probe of the structure of the solar atmosphere. Oscillations in sunspot umbrae \citep[e.g.,][]{1969SoPh....7..351B, 1972SoPh...27...61B, 1972SoPh...27...71G} were reported shortly after the discovery of p-mode oscillations. There are two types of characteristic oscillations in sunspots, the photospheric 5-minute oscillations driven by the p-modes and 3-minute oscillations representing the resonant mode of the sunspot itself \citep[]{1981phss.conf..345T, 2006RSPTA.364..313B, 2015LRSP...12....6K}. 
% you should probably cite these two seminal review articles:
% \bibitem[Khomenko \& Collados(2015)]{2015LRSP...12....6K} Khomenko, E. \& Collados, M.\ 2015, Living Reviews in Solar Physics, 12, 6. doi:10.1007/lrsp-2015-6
% \bibitem[Bogdan \& Judge(2006)]{2006RSPTA.364..313B} Bogdan, T.~J. \& Judge, P.~G.\ 2006, Philosophical Transactions of the Royal Society of London Series A, 364, 313. doi:10.1098/rsta.2005.1701

Studies of sunspot oscillations have been carried out by multiple instruments over the years, including the Solar Optical Telescope onboard Hinode \citep{2007PASJ...59S.631N}, the Nobeyama Radioheliograph \citep{2012ApJ...746..119R}, the Atmospheric Imaging Assembly on board SDO \citep{2012ApJ...746..119R,2014A&A...569A..72S}, the Interface Region Imaging Spectrograph \citep[]{2014ApJ...786..137T, 2015ApJ...798..136Y}, the Goode Solar Telescope (GST) operating at Big Bear Solar Observatory (BBSO) \citep{2013SoPh..288...73M,2015ApJ...798..136Y,2016ApJ...817..117S} and other ground-based telescopes \citep[e.g.,][]{2006ASPC..358..465C, 2016AN....337.1040L, 2019ApJ...882..161A, 2020ApJ...900L..29F}.

%These are the additional references
% \bibitem[Anan et al.(2019)]{2019ApJ...882..161A} Anan, T., Schad, T.~A., Jaeggli, S.~A., et al.\ 2019, \apj, 882, 161. doi:10.3847/1538-4357/ab357f
%\bibitem[Centeno et al.(2006)]{2006ApJ...640.1153C} Centeno, R., Collados, M., \& Trujillo Bueno, J.\ 2006, \apj, 640, 1153. doi:10.1086/500185
%\bibitem[Lohner-Bottcher et al.(2016)]{2016AN....337.1040L} L{\"o}hner-B{\"o}ttcher, J., Bello Gonz{\'a}lez, N., \& Schmidt, W.\ 2016, Astronomische Nachrichten, 337, 1040. doi:10.1002/asna.201612430
%\bibitem[Felipe et al.(2020)]{2020ApJ...900L..29F} Felipe, T., Kuckein, C., Gonz{\'a}lez Manrique, S.~J., et al.\ 2020, \apjl, 900, L29. doi:10.3847/2041-8213/abb1a5

The sunspot oscillations are a specific manifestation of the significant amount of acoustic energy generated through oscillatory motion in near-surface layers. In the upper photosphere, most of the upward propagating waves at lower frequencies (below 5 mHz, or periods longer than 3-4 minutes) are reflected back downward while higher frequency waves continue to propagate upward into the chromosphere. A variety of observations have revealed how energy is deposited in the upper atmosphere due to such wave motions. 
% This paper (Reardon  et  al.,2008) wasn't dealing specifically with umbral waves, but rather the quiet Sun. 
% The issue of shock dissipation is interesting, but I think it poses the bigger need to discuss how the three-minute oscillations observed
% in the sunspot chromospheres often have the sawtooth pattern characteristic of shocks (this goes back to Beckers and Tallant).
% This will likely become important later when discussion the wave signatures
For example, a study by \cite{2008ApJ...683L.207R} using Ca II (854.2 nm) line observations from the Interferometric Bidimensional Spectrometer (IBIS) revealed the presence of a power-law distribution of significant oscillatory power up to 25 mHz, suggestive of the widespread presence of turbulence from shock dissipation in the chromosphere. 
%This can provide a heating mechanism that deposits energy into the chromosphere more evenly than the shock occurrences themselves in both space and time.

Waves also carry information about the structure of the atmosphere in which they propagate. Since the sound speed is a function of temperature in the chromosphere, pressure disturbances with frequencies above the acoustic cut-off frequency can provide direct information about the atmosphere at different heights and the accompanying temperature and velocity perturbations. 

Over many years, numerous efforts have being made in studying the particular oscillatory signals seen in sunspot umbrae. Using H$\alpha$ filters on the 12 inch solar telescope in Culgoora, \cite{1972SoPh...27...71G} was the first to measure sunspot oscillations in H$\alpha$ velocity. Later on, more detail has been revealed using H$\alpha$ line wing data. \cite{1975SoPh...41...71P} showed the intensity variation in red and blue wings of H$\alpha$ at $\pm0.3${\AA} that demonstrate the line-of-sight velocity field as an oscillation on a upward flow.  \cite{1983A&A...123..263U} compared the oscillations measured from several lines (Ca II, Na D1 and D2, Ni I, H$\alpha$) using power spectra analysis and studied the phase relation between them. With the high cadence filtergrams from Universal Birefringent Filter on the Dunn Solar Telescope (DST), \cite{2000A&A...354..305C, 2001A&A...375..617C} studied the relation between running penumbra waves and umbral oscillations in different layers of the solar atmosphere marked by different lines from H$\alpha$ line center and wings. \cite{2007A&A...463.1153T} studied multiple umbral flashes in one sunspot using Ca II and H$\alpha$ intensity images and revealed the coexistence of more than one oscillating mode, suggesting different physical conditions existing in the umbra.

However, studies of wave propagation in the chromosphere are not straightforward. For example, \cite{1976SoPh...49..231M} and \cite{1989A&A...224..245F} found no phase difference between the Doppler shifts of the Ca II infrared triplet lines at 854.2 nm and 849.8 nm, which are formed at different heights, and used this as evidence of a non-propagating component in the chromospheric wave field. However, this conclusion was questioned by \cite{1994chdy.conf...79S}, whose simulations of propagating disturbances predicted little or no phase difference between these lines, similar to the above observations, even though propagating shocks were present. This shows that in interpreting observations it is necessary to consider the radiative transfer effects and the atmospheric structure over the entire range of heights contributing to the observed emissions. 

Previous observations have shown the capability of detecting oscillations in sunspots using radio instruments such as the Nobeyama Radio Observatory \citep{2001ApJ...550.1113S} and the Very Large Array \citep{2002A&A...386..658N}. \cite{2004A&A...419..747L} has demonstrated the feasibility of measuring chromospheric oscillations in the mm range. Such observations have the distinct advantage of directly probing the plasma temperature at different heights in the solar chromosphere depending on the observing frequency.
% Here I would mention some previous observations of sunspot observations in the radio range. Most of those have been at 17-34 GHz with Nobeyama (nothing from BIMA?), but I think they are worth mentioning as the previous observational limits
% \bibitem[Shibasaki(2001)]{2001ApJ...550.1113S} Shibasaki, K.\ 2001, \apj, 550, 1113. doi:10.1086/319820
% -- 17 GHz, compared with SUMER
% \bibitem[Nindos et al.(2002)]{2002A&A...386..658N} Nindos, A., Alissandrakis, C.~E., Gelfreikh, G.~B., et al.\ 2002, \aap, 386, 658. doi:10.1051/0004-6361:20020252
% -- 5-15 GHz, spatially resolved measurements
% \bibitem[Tlatov \& Riehokainen(2008)]{2008A&A...487.1143T} Tlatov, A.~G. \& Riehokainen, A.\ 2008, \aap, 487, 1143. doi:10.1051/0004-6361:20077523
% -- detection of 3 min oscillations at 1.76 cm (no spatial resolution)

% does millimeter emission depend only on local temperature? Or also on magnetic field strength? What is source of mm emission above sunspots? Gyroresonance or bremsstrahlung?
% I think the main source is electron-ion free-free absorption.

Early research \cite[]{1981NASSP.450..263L, 1982A&A...112...16Z, 1984MNRAS.207..731Z} on the oscillations in the chromosphere suggested that the oscillatory modes are produced by a resonant cavity. However, in more recent studies \cite[]{2010ApJ...719..357F, 2015ApJ...808..118C}, a different scenario has been proposed that the 3 minutes oscillation is a direct indication of vertically propagating waves that travel through the gravitationally stratified medium. Despite the growing knowledge and complexity of theoretical models, there remain several open questions related with the fundamental physical mechanisms in the sunspot region. 

% You haven't really mentioned any of the theoretical models for the umbral oscillations...
% I'm writing another model paper regarding the sunspot oscillation, I'm not sure if I should put some material here.
% list several model papers.
The need for better understanding of the fine structure of the 3 min oscillation and its time evolution in sunspots has intensified with the development of better observating tools. Among modern observatories, the Atacama Large Millimeter/submillimeter Array (ALMA) opens up a new era of solar radio observation due to its high spatial and temporal resolution and image quality \citep[e.g.][]{2018A&A...619L...6N, 2019ApJ...881...99M,2021A&A...652A..92N}. When combined with other cutting-edge instruments, such as BBSO/GST, SDO/AIA and IRIS, ALMA can provide unique electron temperature diagnostics that clarify the behavior of the solar chromosphere to propagating waves \citep[see][for a discussion of $p$-mode oscillations seen with ALMA]{2020A&A...634A..86P, 2021A&A...652A..92N}.
% Need a citation about ALMA providing an electron temperature measurement of the umbral chromosphere - SSRv? 
% \bibitem[Loukitcheva et al.(2014)]{2014A&A...561A.133L} Loukitcheva, M., Solanki, S.~K., \& White, S.~M.\ 2014, \aap, 561, A133. doi:10.1051/0004-6361/201321321
% -- used BIMA, 3.5 mm wavelength, discuss discrepancies between models and observations
% \bibitem[Loukitcheva \& Nagnibeda(1999)]{1999ESASP.446..451L} Loukitcheva, M. \& Nagnibeda, V.\ 1999, 8th SOHO Workshop: Plasma Dynamics and Diagnostics in the Solar Transition Region and Corona, 446, 451
% -- earlier work discussion model vs. observation discrepancies, based on 3.4 mm observations

The aim of this paper is to present an analysis of 3-minute oscillations in a sunspot umbra using data gathered by multiple instruments and demonstrate the presence of such oscillatory motion in ALMA Band 3 data. With the help of high-cadence and high-resolution ALMA data as well as the linear dependence between ALMA image intensity and brightness temperature of the plasma \citep{2016SSRv..200....1W}, we report some of the first observations of spatially resolved temperature oscillations in a sunspot umbra.

\section{Observations}

We analysed the oscillation signal measured in the western part of active region NOAA 12470 during  one-hour observing period (18:42~UT -- 19:48~UT on 2015 December 17), which was part of an ALMA solar commissioning campaign whose data are in the public domain. The ALMA dataset we used was provided by the ALMA observatory as part of its Science Verification data release. Previous publications based on this ALMA data release are \citet{2017ApJ...845L..19B}, \citet{2017ApJ...841L..20I}, \citet{2017ApJ...850...35L}, and \citet{2017ApJ...841L...5S}. Only \citet{2017ApJ...841L...5S} used data from the same day (2015 December 17) as considered here, and that study was focused on a small brightening rather than the sunspot itself.  Our observations were conducted using the band 3 receiver (centered at 100 GHz) in a single-pointing (snap-shot) mode. Due to the necessity of phase calibration, blocks of on-target observations were run for approximately 615~s at a cadence of 2~s, followed by a 220-s off-target phase calibrator scan, which resulted in total of five solar scans to be analyzed. In this paper, we will use designations $t1$, $t2$, $t3$, $t4$, and $t5$ to represent each solar scan arranged in time increase order. The ALMA antenna configuration consisted of total 31 antennas with twenty two 12-m antennas and nine 7-m antennas. The longest baseline of the antenna configuration was $\sim$300~m, which resulted in a synthesized beam of 6.3{\arcsec} $\times$ 2.3{\arcsec} \citep{2017ApJ...841L...5S}.

We followed the general calibration method of radio interferometry using the Common Astronomy Software Applications (CASA) package \citep{2012ASPC..461..849P} to create images at 2~s cadence, and found that the result is heavily influenced by phase errors produced by the temporal and spatially variable water content of the Earth's atmosphere, which can be severe enough to render  sunspot oscillations undetectable. In order to counter this effect, we carried out a self-calibration process, which uses the solar signal itself to correct for antenna-based phase and amplitude errors, similar to a method by \cite{2017ApJ...841L...5S}.

% is there a reference anywhere for the self-calibration process? Shimojo? Or is this a new method? In which case I would say "we developed a self-calibration process"
The scheme of self-calibration can be described as follows: 
a) split the data that have been corrected using standard calibration into 10~min sub-datasets $t1$-$t5$ according to the scan index. 
% "treat each solar scan, each with a duration of $\sim$10~minutes, as a separate sub-dataset."
b) For each sub-dataset, make an image from the entire 10-min period using the \textit{CLEAN} task in CASA, to use as a model for self-calibration.  The model is stored in the dateset for later use.  Atmospheric fluctuations will largely average out in such a long integration. 
% by "make a model" do you mean "take the average reconstructed image"?
%yes, this image will be the model to use for later calibration
c) Determine phase corrections for each 2-s time period relative to the model given by step b using the \textit{gaincal} task in CASA, and generate a self-calibration table. 
% could you explain a bit more about how you "determine phase corrections"? You compare the phases at each antenna between the 2-sec and 10-min integration? Or you compare the phases in the Fourier space for the two image? I guess my question is, are the phases determined in the pupil or image plane (which is probably an optical persons way of thinking about it).
% is the self-calibration table generated for each 2-s image?
% the phase correction is calculated using gaincal task, the time-dependent complex gain for each antenna/spwid are determined from the ratio of the data column (raw data), divided by the model column (from step b), for the specified data selection (~2s interval).
% the solution interval is given to be ~2s therefor the self-calibration table is generated for each 2~s image.
d) Apply the phase corrections to each image in the sub-dataset using the \textit{applycal} task in CASA and generate an intermediately corrected sub-dataset. 
e) Iteratively repeat steps b-d, substituting in each intermediate sub-dataset, until no further correction is needed, as determined by checking the calibration table for reduction of the residuals. 
% You had mentioned using this procedure iteratively, but on smaller and smaller sub-blocks of each sub-dataset. Was that done here?
% this method was used in later data set, for the 2015 Dec data, it was done using only ~2s interval setting in gaincal task.

% I checked my script for the 2018/12/20 datasets, the solution interval is set to be ['inf','60','int','int','int'] for each ~10 mins scans, so the calibration is from general to specific 2s images.
% It seems like this procedure will try to drive the individual image to look like the sub-dataset average, right? With the 
% allowable deviations from that average being only what can fall below the level of the residuals? I might be completely
% missing something...
In our case, we applied this phase self-calibration cycle four times, resulting in final residual phase fluctuations of around 10 degrees standard deviation. We applied the scheme to each of the measured 10-min sub-datasets to complete the self-calibration procedure. The result shows a clear improvement in removing the fluctuations caused by the Earth's atmosphere and allows the regular variations due to sunspot oscillations to be measured.

%The field of view (FOV) of ALMA at band 3 (100~GHz) is roughly 60{\arcsec}, which is significantly smaller than the solar disk, therefore the interferometer image does not contain the information from the background level of quiet-Sun emission.

\begin{figure}[ht]
\centering
\includegraphics[width=1\linewidth]{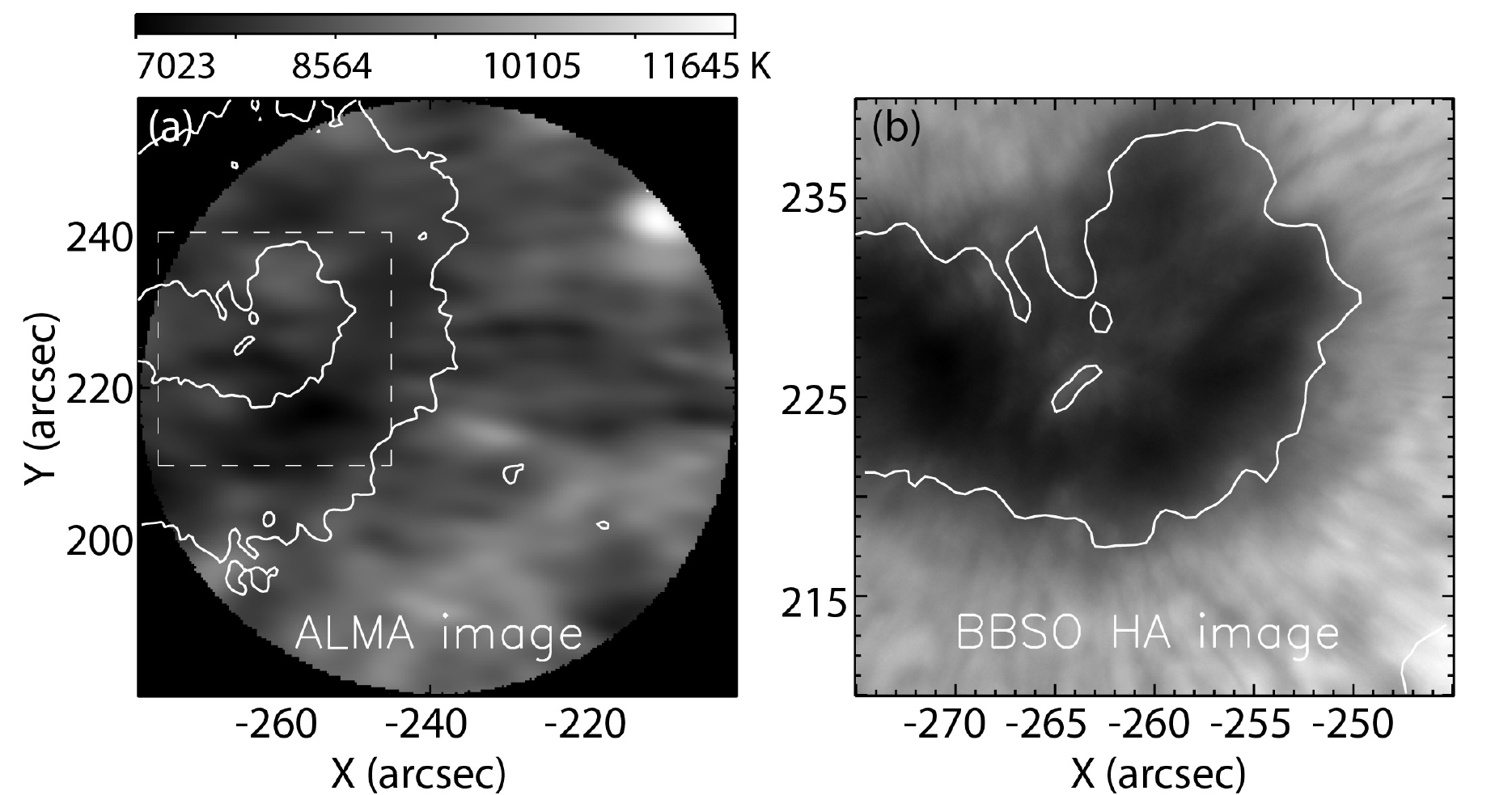}
\caption{(a) ALMA brightness temperature image with contours from HMI continuum image showing the umbral and penumbral boundary of sunspot AR12470 at 19:02 UT. The box marked with a dashed line shows the FOV of the BBSO GST H$\alpha$ image in the right panel. (b) GST H$\alpha$ near blue wing (-0.4 {\AA}) image with the same HMI umbral contour.}
\label{fig:fig1}
\end{figure}

The smallest spatial scale of the antenna configuration is roughly 70{\arcsec}, which is not sufficient to measure the solar background brightness temperature over the band 3 field of view (FOV; $\sim$80{\arcsec} in Figure~\ref{fig:fig1}). To overcome this drawback, the single-dish observations of the full Sun were taken using three other antennas to measure the overall emission close in time to when ALMA was observing the target \citep{2017SoPh..292...88W}. The information from both interferometer and single-dish antennas then were combined using the ``feathering" process in CASA to obtain the true temperature scale on the full-range of spatial scales. The left panel of Figure 1 shows the end result of this full process, which has a pixel scale of 0.3{\arcsec} and FOV as defined by the interferometer images.  The field of view is truncated at the 20\% level of the peak response of the primary (single-dish) beam and has been corrected for the declining antenna response, which becomes more extreme near the edges.
% you never say what was the pixel/spatial scale onto which the ALMA data are interpolated? 0.3"/pixel?
% yes they are 0.3"/pixel

Due to the fact that ALMA is using the equatorial coordinate system (RA-Dec) as the output framework for its data sets, as part of the image processing we performed a coordinate transformation to heliocentric coordinates, to align the ALMA images with those from other solar instruments. In this procedure, the center of the field of view (FOV) of each solar image in one scan was transformed based on the heliocentric coordinates at the mid-time of the scan to align all of the images to a fixed time. A time dependent rotation was applied to take into account the rotation between these two frames, leaving the images aligned with solar north.

%the rest may not be needed
%a special procedure is required to align it with results from the solar instruments, in which the helio-centric coordinate system is used. The transformation includes shifting the ALMA center vector to align with the helio-centric coordinate system, and rotating the ALMA image according to its new center, to align with solar north. The first step is based on a rotation matrix generated using RA and Dec information of the ALMA image center pixel and the Sun as well as solar P-angle (position angle of solar north) at that time, while the second step only involves a rotation by the solar P-angle. The coordinate transformation was done using IDL procedures.

To compare the ALMA images with those at other wavelengths, we obtained the UV and EUV intensity data cube from the Atmospheric Imaging Assembly (AIA) on board the Solar Dynamics Observatory (SDO) \citep{2012SoPh..275...17L} in the bandpass of 171~{\AA} (Fe IX: the quiet corona) and 304{\AA} (He II: the chromosphere and transition region). The time cadence of these two data sets is 12~s and the pixel resolution is 0.6{\arcsec}. The duration of the AIA data we used covers the full ALMA solar observation from 18:42~UT to 19:48~UT. Images of both wavelengths were extracted from the full-disk data to fit the FOV of ALMA and the center coordinates of these images were shifted to correct for solar rotation. The continuum intensity data product from by the Helioseismic and Magnetic Imager (HMI) on SDO was used for alignment of between the SDO and GST images.

The Goode Solar Telescope (GST) operating at BBSO  \citep{2010AN....331..636C} observed this AR nearly simultaneously with ALMA (from 18:50~UT to 20:59~UT) at H$\alpha$ line-center and off-band wavelengths ($\pm0.4$~{\AA} and $\pm0.8$~{\AA}) using the Visible Imaging Spectrograph (VIS) instrument. The spatial scale of the GST H$\alpha$ images is 0.034{\arcsec} per pixel, the VIS bandwidth is 0.08~{\AA}, and the time cadence was 40~s. Speckle reconstruction is normally performed by taking 100 frames in rapid succession and applying the algorithm to obtain a single sharper image. However, this algorithm can produce occasional glitches in low-light regions such as sunspot umbrae, which are the focus of our study. Since high spatial resolution is not needed for large-scale features of umbral oscillations, the unreconstructed H$\alpha$ images were used. The GST observations cover the entire ALMA observing period $t2$--$t5$ but are missing the first few minutes during $t1$ set. Since our interest is on the joint ALMA-GST coverage, we mainly focus on analysing the $t2$--$t5$ ALMA sub-datasets.

We also used the slit-jaw images in Mg II k band (2796~{\AA}) from the Interface Region Imaging Spectrograph (IRIS) which is formed in the chromosphere \citep{2014SoPh..289.2733D} as a comparison to the ALMA temperature oscillation. IRIS started observations during $t3$ and suffered some cosmic-ray noise during the first 10 minutes, which can be seen later in the time-distance plots.

% No mention of the IRIS data, which is shown in Figure 2?
% How was ALMA co-aligned with AIA or the other wavelengths?

\section{Analysis}
\begin{figure*}[ht]
\centering
\includegraphics[width=0.9\linewidth]{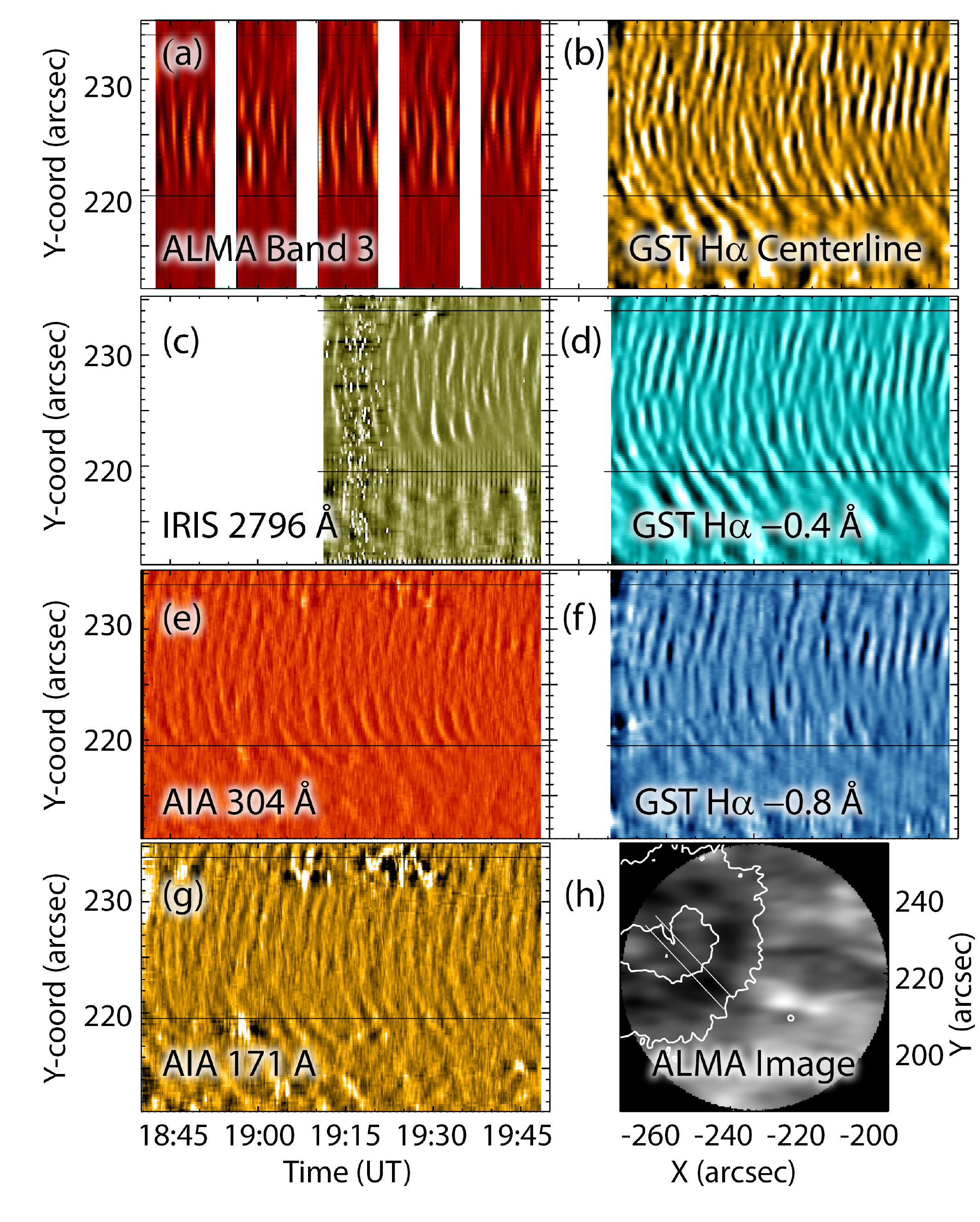}
\caption{Time-distance plots based on the full duration of ALMA band 3 observation, 18:42:33-19:48:31 UT. An average of pixels between the two diagonal lines marked in panel (h) is used to constructed each time-distance plot. From (a) to (g): ALMA Band 3; BBSO GST H$\alpha$ center line; IRIS 2796{\AA}; BBSO GST H$\alpha$ near blue wing (-0.4 {\AA}); AIA 304{\AA}; BBSO GST H$\alpha$ far blue wing (-0.8 {\AA});  AIA 171{\AA}. The horizontal solid lines at y-offsets 219\arcsec\ and 234\arcsec\ mark the umbral boundaries, and the gaps between each ALMA band 3 time block are represented with white gaps. (h): ALMA image of sunspot NOAA 12470 with the boundaries of the slit and the contours from HMI image overlaid to show the umbral boundary.}
\label{fig:fig2}
\end{figure*}

To visualize the oscillatory motion in the sunspot in various wavelengths, we show in Figure~\ref{fig:fig2} time-distance plots along a line (Figure~\ref{fig:fig2}h) that crosses both umbra and penumbra. The double solid lines in Figure~\ref{fig:fig2}h mark the width of the slit over which the data are summed, with an ALMA map for reference, overlaid with HMI contours that mark the boundaries of the umbra and penumbra. The time-distance plots were constructed for each wavelength after co-alignment, as shown in the other panels of Figure~\ref{fig:fig2}.  The $y$ coordinate in each plot corresponds to the $y$ coordinate of the center of the slit in Figure~\ref{fig:fig2}h. Note that the plots are marked with the $y$ coordinate, not distance along the slit.  Multiply the $y$ coordinate by $\sqrt{2}$ for this roughly $45^\circ$ slit to convert to distance.
% so the y-axis in the figure 2 plots is not _actually_ a physical distance (with a conversion of 1 arcsec = 720 km), 
% but rather the value of the y-coordinate along the line. Because the line is inclined at about 45 degrees, the 
% physical distance is ~sqrt(2) times greater. So the black lines in the plots are at 220 and 234 arcsec as the 
% y-coordinate, which actually means a distance on the Sun of 14 arcsec * 720 km/arcsec * 1.4 = 14 Mm (not 10 Mm).
% yes the helioprojective system does have this effect but I doubt if it is ~45 degree since the whole sun is about +- 1000 arcsec.
% The ~45 degrees is the angle of the line in Figure 2(h) with respect to the helioprojective coordinates. The length in the Y direction is (235-215)=20 arcsec, but the traced line is actually 35 arcsec long.

The horizontal black lines in each panel mark the extent of the umbra. The ALMA data for $t1$-$t5$ are in Figure~\ref{fig:fig2}a, with gaps indicating times of no data due to calibrations. GST H$\alpha$ center-line and blue-wing data are shown on the right (Figs~\ref{fig:fig2}b,d,f). IRIS 2796~\AA\ slit-jaw data are shown in Figure~\ref{fig:fig2}c. AIA 304~\AA\ and 171~\AA\ data are shown in Figs~\ref{fig:fig2}e,f.  Non-uniform chevron-shaped features \citep{2006SoPh..238..231K} that tilt in the same direction can be seen in Figs~\ref{fig:fig2}b--g, suggesting an outward propagation towards the penumbra. The 3-minute oscillatory motion in these wavelengths also extends somehow into the penumbra (above or below the solid black lines in Figure~\ref{fig:fig2}a--g), but with some merging and increased separation to match the penumbral 5-minute oscillation period. This signature of propagation seems to be far less apparent in the ALMA data, and the oscillatory signature is only prominent in the umbra.
% the IRIS data don't show much in the way of penumbral waves either, and it's not very clear in the 304 Å or 171 Å channels.

%adding panel mark to figure

%The time-distance plots in Figure~\ref{fig:fig2} show that the 3-minute oscillations are spatially constrained within the umbra for every wavelength presented in the figure, while some wavelengths (especially Figure~\ref{fig:fig2}b,d) show the longer-period 5-minute oscillations in the penumbra (below the 219\arcsec\ umbral boundary). The oscillations are disturbances traveling from the center to the edge of the umbra as shown by the chevron-like shape of the oscillation patterns seen in most wavelengths.  However, unlike the other wavelengths this traveling wave pattern in not so obvious in our ALMA data, although that may be due to the relatively low spatial resolution. The ALMA data also shows neither strong 5-minutes oscillation in the penumbra nor any sign of transitions of waves beyond umbra. 

%A possible suggestion is that there may exist a cavity within the formation height of ALMA band 3 wavelength that leads to a behaviour of standing wave. Meanwhile the absent of 5-minute oscillation signal in ALMA band 3 may also suggest that the 5-minutes p-mode wave in penumbra can not create a corresponding temperature variation signature that can be captured by ALMA band 3. 

%should mention that the duration of ALMA scans also limited the finding of 5 minutes oscillations

\begin{figure*}[ht]
\centering
\includegraphics[width=0.9\linewidth]{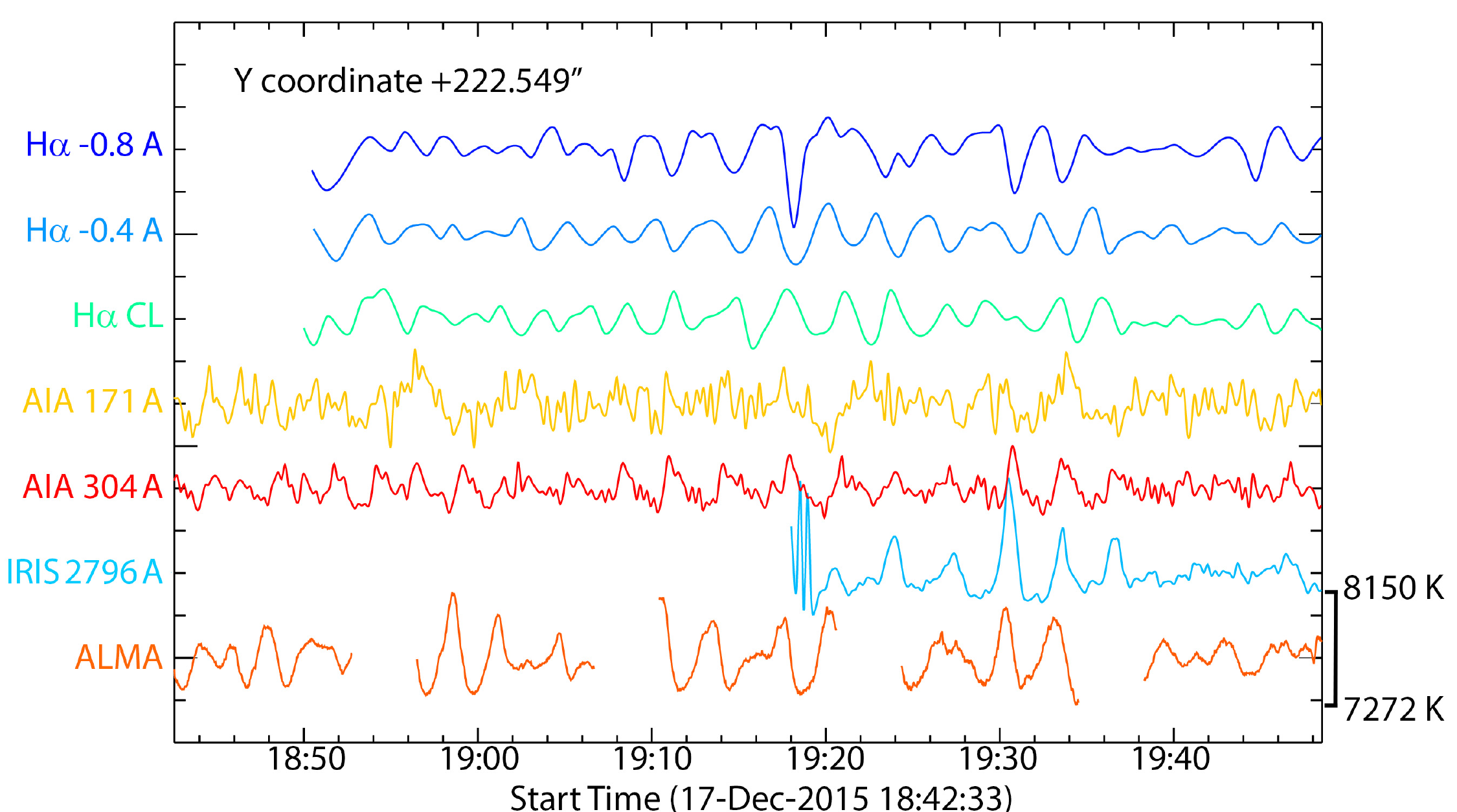}
\caption{Light curves from an average over pixels between the two slit in Figure~\ref{fig:fig2}h at $x=-251.0$\arcsec, $y=222.5$\arcsec. The point along the slit was chosen due to its good representation of the wave pattern. The name of instruments and observation band are on the left hand label, and the ALMA temperature range is shown with the scale on the right.}
\label{fig:fig3}
\end{figure*}

%Although limited by the duration of each scan, ALMA is not capable of detecting longer than 5-minute period oscillations, the 
Our discovery of umbral oscillation in ALMA band 3 is groundbreaking in many respects. ALMA data provide a unique probe of temperature changes in the solar atmosphere in a relatively thin layer where the opacity is near unity \citep{2015A&A...575A..15L}.
% but remember, that layer is not fixed at a constant geometrical height, which can make interpretation of the temporal
% behavior of that diagnostic somewhat problematic.
It is of interest to compare these oscillations in temperature with intensity oscillations provided by other instruments that probe different layers of solar atmosphere. Figure~\ref{fig:fig3} compares the light curves obtained from a cut at $x=-251.0$\arcsec, $y=222.5{\arcsec}$ in the time-distance plot (Figure~\ref{fig:fig2}) for multiple instruments. The location of the cut was chosen based on the wave patterns in all wavelengths shown in Figure~\ref{fig:fig2} so that it covers the region of strongest oscillations. The data have been arbitrarily scaled and offset for better comparison. Though separated by gaps, we can still see some correspondence between light curves of different wavelengths, which enable the calculation of phase relations that will be discussed later.

%To further reveal the nature of the oscillatory signals and study the connections between signals from different position of the sunspot as well as different bands, we applied Fourier analysis to the ALMA and BBSO data. The 3-minutes umbral-chromospheric oscillation signal was verified by the power spectrum from both data sets. Although the temporal resolution of ALMA ($\sim$2 s) and BBSO ($\sim$40 s) data are different, they are both well below the Nyquist limit (90~s) of 3-minutes oscillation, therefore capable of showing the 3-minutes oscillations in the power spectrum.

%change the location of the Ireland's paper citation
In order to study the oscillation power, we applied the Fourier power spectrum for each pixel in both the ALMA map and GST image time series. Although the temporal resolution of ALMA (2 s) and GST ($\sim$40 s) data are quite different, they both contain sufficient samples to resolve 3-minute oscillations and therefore can be used to measure the 3-minute oscillation power. The power spectrum was calculated as follows. For each ALMA (10-minute) solar scan, a map cube structure was made according to the method described above so the time series of each pixel can be easily obtained. The GST data were selected based on start and end time of each ALMA solar scan and made into map cube as well.
%Due to the different time cadence of ALMA and BBSO, as well as the $\sim$200 s gaps between ALMA scans, the BBSO data were cubic-spline-interpolated to the same times as ALMA images in each of the $t1$-$t5$ data cubes to achieve the same sampling rate as well as observing duration, in which case the power spectrum from both data sets are now comparable. 
% This seems to imply that the BBSO data were broken up into similar 10 minute blocks corresponding to the time of the ALMA scans
% But figure 3 shows the H-alpha data interpolated continuously over the full observing interval?
% So maybe the H-alpha data were interpolated onto a 2-second time grid, and then the same observing intervals as ALMA were selected out?
%
% Doing FFT's on the same observing intervals is the right thing to do to compare power maps and phases. 
% But it seems unnecessary to upsample the BBSO data by a factor of 20 in the temporal axis. But it shouldn't affect the results.
Inspired by the method used in \cite{2015ApJ...798....1I}, the time series of each pixel in the ALMA and GST data cubes were obtained and normalized by $(I(t))-\langle I(t) \rangle)/\langle I(t) \rangle $, {where $I(t)$ is the varying brightness temperature for ALMA or varying intensity for GST, and $\langle I(t) \rangle$ is the mean over each 10-minute ALMA scan.} 
% is the intensity average <I(t)> taken over one solar scan (t1-t5)? Or the average over all the scans together?
% I'm guessing the former...
%
% yes, the average is taken per solar scan.
% By the way, Figure 3 doesn't seem to be cited in the paper text anywhere. Maybe here?
% Figure 3 is cited at previous paragraph, 
% KPR - oops, missed it.

\begin{figure*}[ht]
\centering
\includegraphics[width=0.8\linewidth]{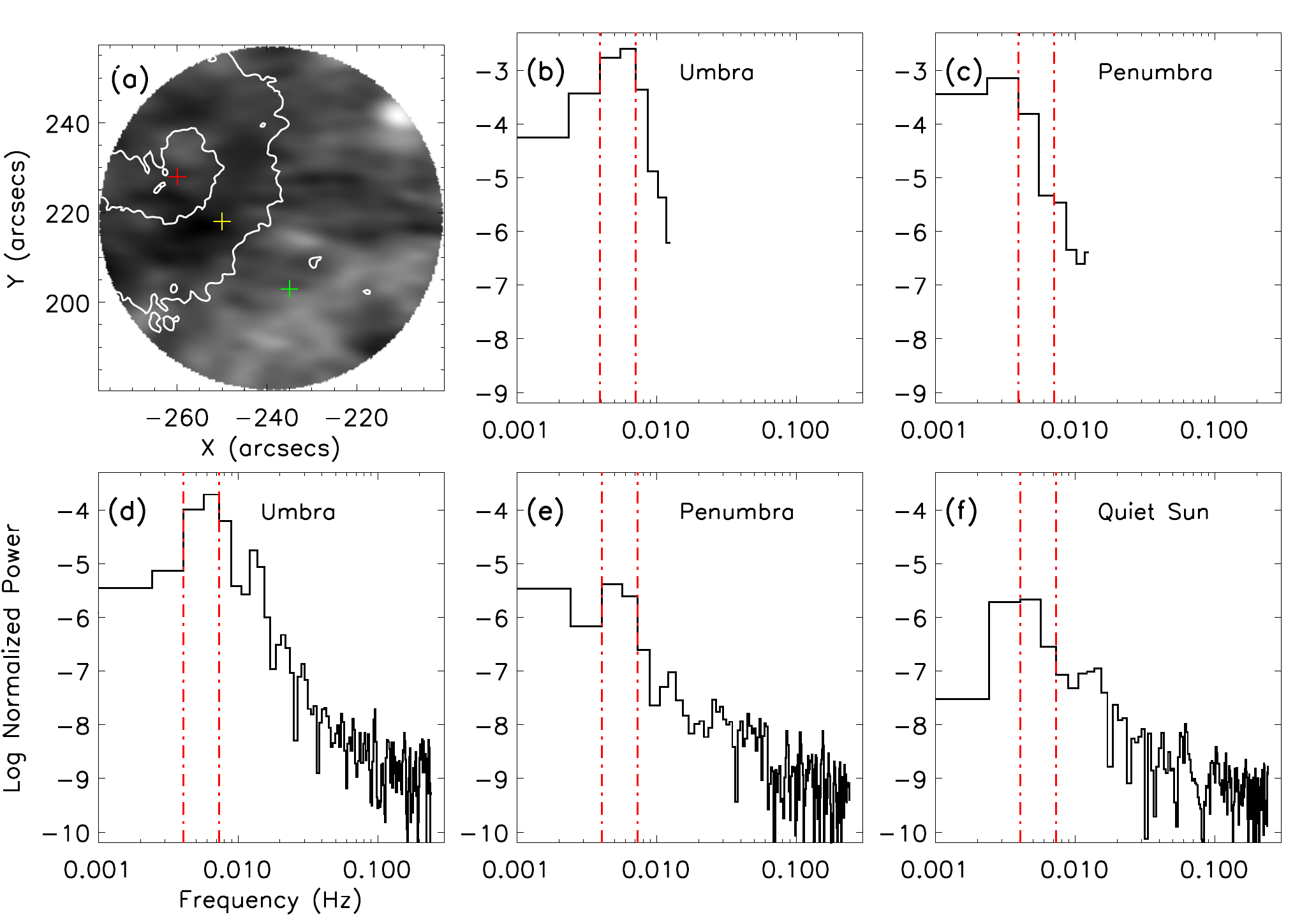}
\caption{Three examples showing the power spectrum in the sunspot umbra, penumbra and quiet Sun regions. (a) ALMA image for locating the sample points, marked with plus signs. The inner white contour is the boundary of the sunspot umbra while the outer contour is the boundary for the sunspot penumbra. {(b) Power spectrum of GST H$\alpha$ near blue wing (-0.4 \AA) data from red point in the sunspot umbra showing a strong peak in the 3-minute oscillation band marked with dot-dashed line. (c) The same as (b) for yellow point in the sunspot penumbra. (d) Power spectrum of ALMA data from the red point in the sunspot umbra. (e) (f) the same as (d) for sunspot penumbra and quiet Sun region.}}
\label{fig:fig4}
\end{figure*}

A Hanning window was applied to the normalized time series to minimize edge effects. The full spatial resolution was retained during the process in order to reach the finest possible detail on the distribution of frequency-related physical contents as a function of location. 
% the full spatial resolution of the BBSO or ALMA data? Maybe specify or give the pixel scale here?
% I add the ALMA spatial resolution info in the observation part, the BBSO spatial resolution is in that part as well.
We then applied the Fourier transform to the processed time series to form the power spectrum from each pixel. Figure~\ref{fig:fig4} shows examples from three pixels in {both the} ALMA {and GST H$\alpha$ near blue wing (-0.4 \AA)} data cubes for time $t2$, one in the region of umbral oscillations, one in the penumbra, and one in the surrounding quiet Sun. {No GST power spectrum could be obtained for the quiet Sun due to limited spatial coverage (see Figure~\ref{fig:fig1}b).}

%change sample points in figure 4 so that all points are in the slit region in figure 2
%A glimpse of the result from such procedure can be find in Figure 4. We chose three points from ALMA $t2$ power spectrum map within the region marked by the slits showed in Figure 2(h) to make a quick comparison between different locations: sunspot umbra, penumbra and quiet sun. These sample points are marked in plus signs in the ALMA map in Figure 4 top left panel with white contours shows the boundary of the sunspot. The detailed helio-centric coordinate is also given in sub-titles of the individual power spectrum.

The differences in shape among these power spectra are consistent with what was seen in the ALMA time-distance plot in Figure~\ref{fig:fig2}a. 
% The quiet-sun point lies out side of the range covered in the time-distance plot
The 3-minute band of the power spectra, which is {the two closest bins to $0.0056$ Hz} marked with red dot-dashed lines, is especially strong for the umbral power spectrum. From Figure~\ref{fig:fig4}, we see that the peak of the umbral power spectrum is one to two magnitudes larger than the others {in both data sets.}
%but there is also a steeper high frequency tail represented by the fitted blue dashed line in the figure. 
% KR - Is the steepness of the tail just defined by the connection between the (spatially variable) three-minute power peak 
% and a relatively constant noise floor at around log(P) ~ -9 ?
% it is formed using 1-D fitting
The umbral spectrum {of ALMA data} also shows some evidence for harmonics {near $0.015$ Hz}, which strongly suggests that the oscillation deviates from a sinusoid, as we demonstrate later.  
% are these harmonics the peaks at 0.015 and 0.02 Hz?
The frequency distribution of power is similar for the penumbra and quiet Sun. 
%also may need to mention the duration of each ALMA scan may cause the lost of 5 mins oscillation signal.
% KR - I wonder if the self-calibration technique could be causing the attenuation of signals whose period is close to that of the observing block duration?

%tian2014[IRIS]tian2014the Interface Region Imaging Spectrograph \citep[IRIS][]{tian2014, 2015ApJ...798..136Y}. 
%bresultsobserved theaation from toiison ALMA observations fromaseach solar scan tookand wassorderarray s in total22 9 during the solar campaign observation.is- theirregular motioncausedanySs-tend tosplitthedata onwhilesuch combination samemostfiltergram datagathered also, described next alsoactive region nein BBSO6, with a - time gap between each imageof GST images   of umbral oscillations the broad for further analysisrsation smissed . Based of ALMA and BBSO, thes ofof the  we show acquireagrees

\begin{figure}[ht]
\centering
\includegraphics[width=1.0\linewidth]{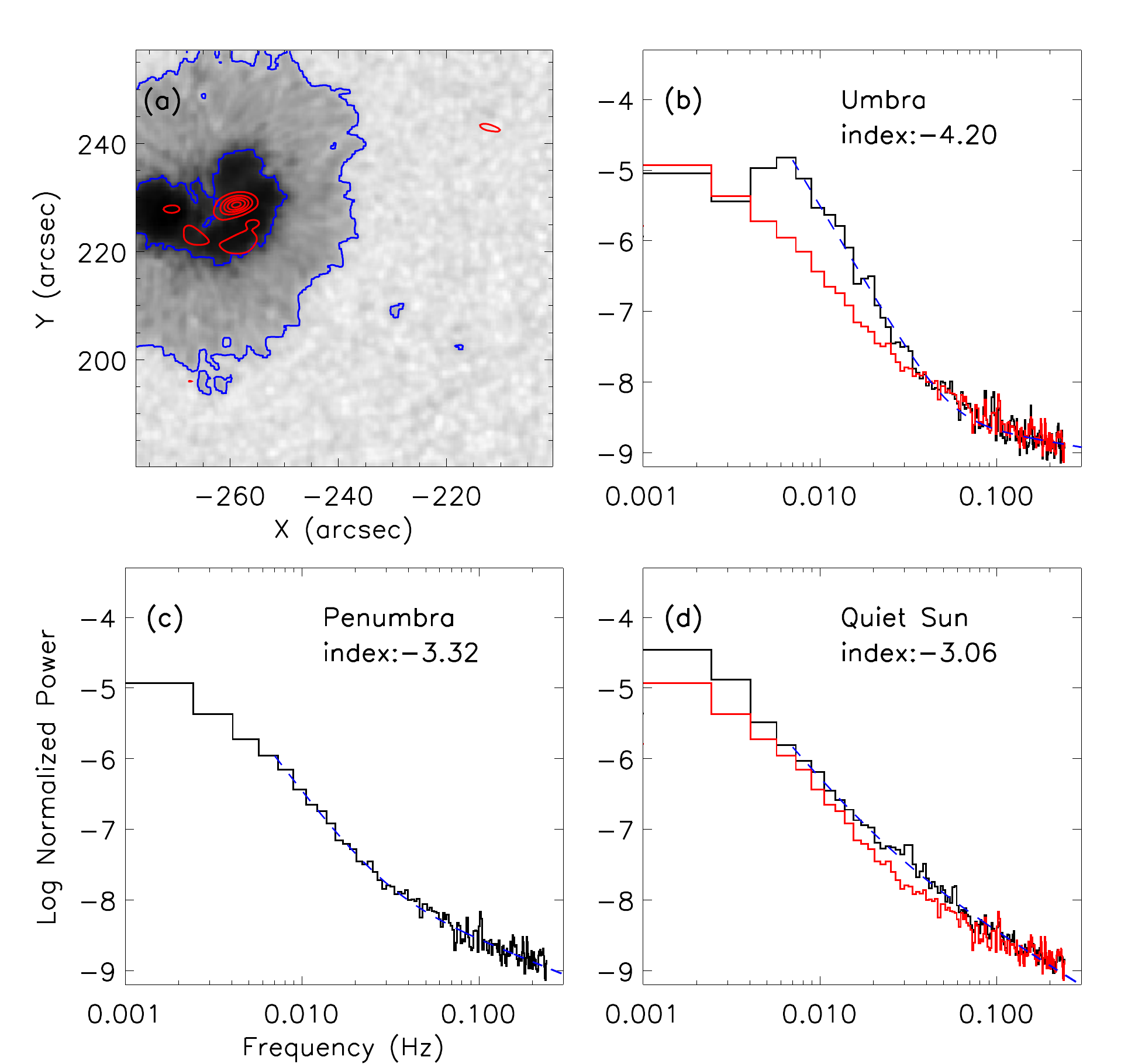}
\caption{{Comparison of ALMA power spectra averaged over umbral, penumbral, and quiet Sun regions. (a) HMI continuum image at 19:02 UT, overlaid with red contours showing the 3-min oscillation power and blue contours outlining the umbra and penumbra from the HMI continuum image. (b) Averaged umbral power spectrum with blue dashed line showing a region well fit by a power-law. (c) Same as (b) for the penumbral region. (d) the same as (b) for quiet Sun. The averaged penumbra power spectrum in panel (c) is over-plotted in red in both (b) and (d).}}
%Left panel: HMI continuum image at 19:02 UT, overlaid with red contours showing the 3-min oscillation power and blue contours outlining the umbra and penumbra from the HMI continuum image. Right panel: ALMA power-law index map generated with $t2$ data, with the same contours from the left panel overlaid.
\label{fig:fig5}
\end{figure}

%Since 3-minutes oscillation band is the dominant component in the power spectrum, it is important to study the geometric information regarding the distribution of the 3-minutes oscillation power in umbra. 
To characterize the power spectrum spatial distribution, for each pixel we calculated the integrated power in the 3-minute band {marked with the red dash-dot lines in Figure~\ref{fig:fig4}}. 
This procedure was carried out for each 10-minute ALMA scan $t2$ to $t5$.  The result for $t2$ is shown in Figure~\ref{fig:fig5}, where the 3-minute oscillation power is shown with red contours. Blue contours outline the umbral and penumbral boundary as seen in the HMI continuum image in panel (a).
% KR - what about showing the actual map of three-minute power (with the same contours), rather than the whitelight image? 
% the 3-minute power map would only show several strong region therefore it is hard to tell the geometry relation between 3-minute power region and sunspot umbra.
%ALMA band 3 and BBSO H$\alpha$ multi-band data set so we can form pictures of the 3-minutes power distribution in both wavelength and time domain. Furthermore, since Figure 4 revealed some connection between the power law index and the peak magnitude of the power spectrum, it is also necessary to study the distribution of power spectrum index.

%Figure 5 left panel shows the 3-minutes power contours (red ones) of the ALMA $t2$ power spectrum map, overlaid on the HMI white light image. The sunspot inner and outer boundaries are highlighted using blue contours. 
We find that the strong 3-minute oscillation power falls solely inside the umbral boundary. The oscillations seem to be strongest in the western part of the umbra, while in the eastern region separated by a small light bridge structure, the 3-minute oscillations are present but weaker. This conclusion may be affected by the declining ALMA sensitivity at the edge of the primary beam, however.
%in the middle of the sunspot also separate the contours into two parts, and the west part has more intense 3-minutes power than the east part.

{To further compare the power spectra in different regions, we calculated the mean power for each frequency bin using all the pixels in the umbra, penumbra and quiet Sun to create the averaged power spectrum of ALMA t2 data for each region. A combination of two power-law functions, $af^{b}+cf^{d}$, was used in fitting the power-law tail starting from the frequency bin immediately after the 3-minute range. The power-law index of the lower frequency term is shown in Figure~\ref{fig:fig5} panel (b) - (d). The averaged penumbral power spectrum was over-plotted using red lines on the umbral and quiet spectra for inter-comparison. The steeper slope shown in panel (b) by the blue dashed line reveals the influence of the high 3-minute power in umbral region, while the power-law indexes from the penumbra and quiet Sun are less steep. It is interesting that the umbral and penumbral high-frequency tail are essentially the same above 0.03~Hz (panel b), while the quiet Sun spectrum in panel (d) shows a notable excess near 0.03~Hz that contributes to a shallower and extended low-frequency slope. A red-noise spectrum above about 0.07~Hz is the same for all three regions, and may reflect residual fluctuations due to the Earth's atmosphere that have not been removed by the self-calibration procedure. Note that although the 5-minute oscillations cannot be well-resolved in the frequency domain, there is no sign of a peak near 0.0033~Hz in the penumbral spectrum of Figure~\ref{fig:fig5}c, which further confirms what we learned from the time-distance plots---that the penumbral 5-minute oscillations seen in H$\alpha$ (Figure~\ref{fig:fig2}b,d) are not apparent in the ALMA band 3 data (Figure~\ref{fig:fig2}a).}

%We notice that the power-law index map in the right panel of Figure~\ref{fig:fig5} seems to show slightly shallower slopes on average in the penumbra, while the umbra has regions of steeper slopes, consistent with the trend in Figure~\ref{fig:fig4}. In this map, we find that the 3-minute power contours are separated by a region of shallower slopes that shares a similar shape and location as the light bridge in the HMI white light image, while the  region of steeper (i.e. lower) slopes the index map inside the sunspot umbra is partially but not completely overlaid by the 3-minute power contours. 
%Since brighter region in the index map indicate gradual slope of the power law fitting, a possible assumption is that the light bridge structure some how serve as wall that block the 3-minutes signal from propagating towards the east part of the umbral region.

\begin{figure*}[ht]
\centering
\includegraphics[width=0.9\linewidth]{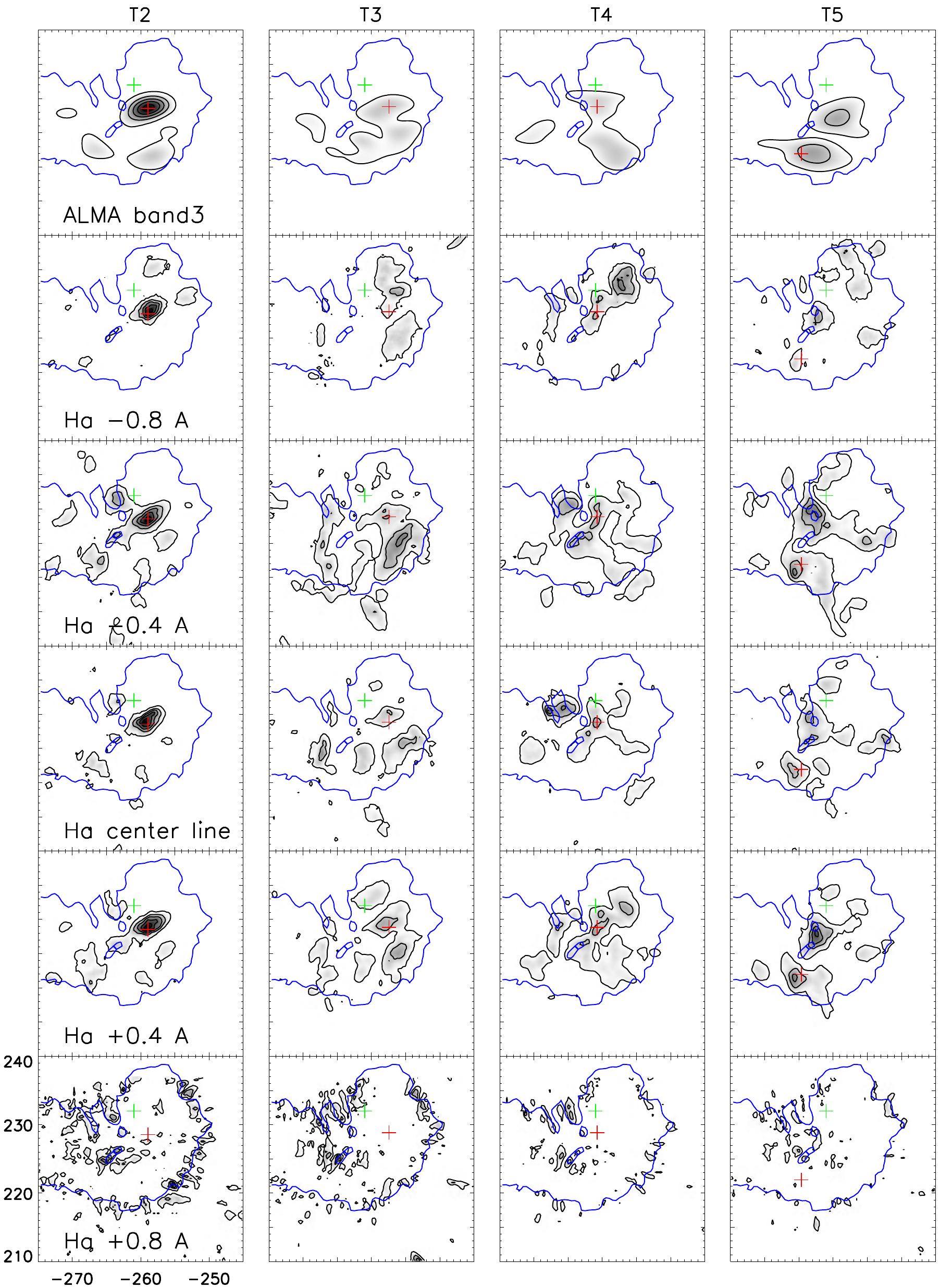}
\caption{Spatial distribution and relative power level of 3-min oscillations in various bands. For this comparison the H$\alpha$ data have been temporally interpolated and spatially smoothed to match the ALMA resolution. The contours in ALMA and the H$\alpha$ sub-band images represent 10, 30, 50, 70 and 90 percent of the peak value of the 3-min oscillation power in each wavelength. Red and green plus symbols mark regions of high and low power, respectively, whose light curves are plotted in Figure~\ref{fig:fig7}.
%The H$\alpha$ far red wing (+0.8\AA) data has the smallest peak value of the 3-min oscillation power among all H$\alpha$ sub-bands, (from H$\alpha$ -0.8\AA to + 0.8\AA: $8.16 \times 10^{-7}$, $4.28 \times 10^{-6}$, $1.93 \times 10^{-6}$, $4.27 \times 10^{-6}$, $1.6 \times 10^{-7}$), and shows less 3-minutes oscillation signal in umbra compared with the rest data sets.
}
\label{fig:fig6}
\end{figure*}

Figure~\ref{fig:fig6} shows the distribution of 3-min oscillation power in the $t2$ to $t5$ time ranges (columns) and all bands from ALMA and GST (rows). The rows represent the spatial evolution of the 3-min power for the wavelength band marked in the first column. The sunspot umbra is outlined in blue (see Figure~\ref{fig:fig5}) while the black contours with gray shading show the 3-min power maps of each data set. For a given band, the contour levels were set to be the same for $t2$ to $t5$, which are 10, 30, 50, 70 and 90 percent of the maximum value of oscillation power over the four power maps. 
%These maximum values are: $2.81 \times 10^{-7}$, $8.16 \times 10^{-7}$, $4.28 \times 10^{-6}$, $1.93 \times 10^{-6}$, $4.27 \times 10^{-6}$, $1.6 \times 10^{-7}$ respectively.

From Figure~\ref{fig:fig6} we see that the shape of {GST} H$\alpha$ far red wing (+0.8\AA) 3-min power map contours is less organized and weaker than the others. While the other contours are generally constrained to lie well within the umbra, the power in the H$\alpha$ far-red wing mainly lies along the umbral boundary. The relative amplitude of the +0.8\AA\ power is also 5-50 times lower than in other bands, and hence we do not include it in our subsequent analysis. 
%abnormal may be explained by the low 3-minutes oscillation power value listed before. As we compare all peak values of H$\alpha$ data, the far red wing one is much smaller than the far blue wing (-0.8\AA) one, thus may leads to a suggestion that the SNR in the original far red wing data set is worse than the others, therefore should not be used in the study of the 3-minutes umbral oscillation.

For the remaining bands, the 3-min power was in all cases strongest and most concentrated in $t2$ and reduced to 20-40\% of that value in $t3$ to $t5$. Although it is difficult to tell from the shape and location of these contours, which suffer from the limited cadence of the H$\alpha$ data, the overall pattern is consistent with the often seen appearance of new centers of oscillation power plus the lateral spreading of wave energy from previous centers \citep{2020ApJ...896..150Y}.
%cite Yurchyshyn 2020 paper 
%in all band also has more similarity in $t2$ column compared with the rest ones. When the 3-minutes oscillation power is weak, as seen in $t3$-$t5$, we find little to no connection between the location of the peak value as well as the morphology of the contours. During that period, we can also see that the contours turn to expand when compared with $t2$, and can be also seen crossing the light bridge region in ALMA band 3 data. All these phenomenon suggesting that the cavity as well as the light bridge structure only appears to constrain the waves that mainly composed by 3-minutes component, and when the energy is diffused into other oscillation band, we will see the oscillation signal extend outside the cavity.

\begin{figure*}[ht]
\centering
\includegraphics[width=0.8\linewidth]{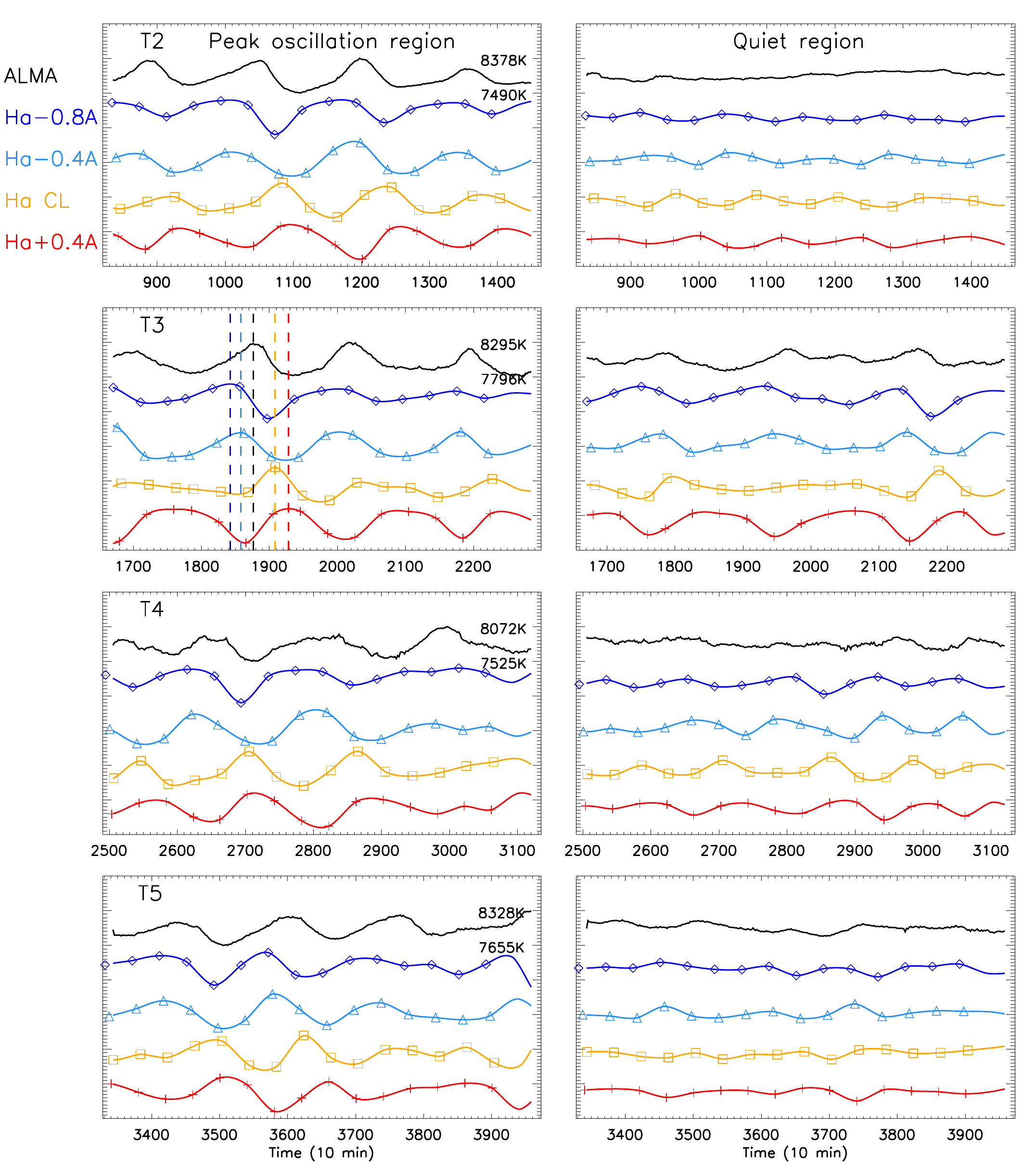}
\caption{A comparison of light curves from pixels of strong oscillation (left column) and those from a relatively quiet region (right column) from locations marked with red and green plus sign in Figure~\ref{fig:fig6}. Different bands of ALMA and GST H$\alpha$ data, which follow the same order top-to-bottom as in Figure~\ref{fig:fig6}, are plotted with different colors as shown by the labels adjacent to the upper-left panel. The time axis is marked in seconds, sequentially from the start time of the ALMA observations at 18:42:33UT.
%The H$\alpha$ far red wing (+0.8\AA) is not shown due to the lack of 3-min oscillation power. 
The dashed lines in $t3$ left column mark the peak times of these light curves for one particular oscillation, as discussed in the text. The ALMA maximum and minimum brightness temperatures are also shown in each left-hand panel, showing a fluctuation of $\pm250$-500 K.}
\label{fig:fig7}
\end{figure*}

To determine the phase relationships among ALMA band 3 and GST H$\alpha$ sub-bands, we plot in Figure~\ref{fig:fig7} the light curves from these wavelengths taken in a region of high oscillation power (red plus symbol in Figure~\ref{fig:fig6}) and for comparison a neighboring region of low power (green plus symbol in Figure~\ref{fig:fig6}).  The selected locations for the high power sample are different from $t2$ to $t5$, but within each given time block, the same position is used for every wavelength. The ALMA 2-s cadence data are plotted directly (black curve), while the lower-cadence H$\alpha$ measurements are plotted in different symbols and colors, with the solid curves through the points showing the cubic spline interpolation to the ALMA cadence.
%In addition to the interpolated BBSO data, the non-interpolated point is also plotted on the light curves in the figure represent by different symbols.

To better reveal the phase relationships among the different bands shown in Figure~\ref{fig:fig7}, we show five vertical dashed lines in the left column T3 panel that indicate the peak positions of the five light curves for one particular oscillation. The colors match the colors of light curves for the corresponding wavelengths. These dashed lines reveal a general sequence of the phase relationship, with H$\alpha$ far blue wing (-0.8 \AA\ or -0.08 nm) peaking first, then in order of peak time the H$\alpha$ near blue wing (-0.4 \AA), ALMA band 3 (100 GHz or 3.0 mm), H$\alpha$ center line and H$\alpha$ near red wing (+0.4 \AA). We calculate the delay times based on these peak positions, relative to the H$\alpha$ far blue wing to be 14 s, 32 s, 64 s and 82 s. This timing pattern is followed remarkably faithfully in the other individual oscillations in the left column, with similar time lags. Since the locations of the sample point in the umbra change in the different periods $t2$ to $t5$, the relatively stable phase shift between these bands suggests that it is a characteristic of the propagating waves that should be accounted for in any physical interpretation.
%Given the formation height of ALMA band 3 and H$\alpha$ sub-bands, we can estimate the phase speed of the vertical propagating wave. However, based on the study done by Loukitcheva \citep{2015A&A...575A..15L}, the effective formation height of ALMA band 3 (3 mm) can very from $\sim$ 1000 km (Fontenla et al. 2009) to $\sim$ 1860 (Maltby et al. 1986), make it very hard to calculate a solid number for the phase speed. 

%need more detail about formation height of ALMA band3 and H alpha. From Loukitcheva's paper, the formation height is estimate to be 1200km to 1500km, while the formation height of H alpha is more complicated and depends on both Ne and temperature(J. Leenaarts, 2012). Thus a certain number that represent the phase speed may not be give under such variety

We also compared the light curves from the strong 3-min oscillation region in the left column with the ones from a nearby, relatively quiet umbral region (green plus symbols in each panel of Figure~\ref{fig:fig6}) in the right column for $t2$-$t5$. Unlike the case for the peak oscillation region, where the location is different for each time window, the quiet region selection is fixed for all of the time windows. For each time window, the amplitude range was scaled to match the left column to show the difference between these two regions. One can see that for $t2$, $t4$, and $t5$ ALMA temperature oscillations are not apparent when compared to the high-power region, although velocity fluctuations in the H$\alpha$ bands still persist at a reduced level. In $t3$ the ALMA oscillations do appear in the quiet region, again at a reduced level.
% I'm not sure that the light curves is the best (or even proper) way of highlighting differences in the phase delays. Especially if the weaker oscillatory signal is scaled like the stronger signal, effectively making it very difficult to see the peaks and valleys.
% Is there any way to do a numerical determination of phase (for FFT power spectra)? 
The significant differences in oscillatory power across the umbra, both in temperature and velocity diagnostics, indicate rapid spatial and temporal modification of the wave propagation characteristics.
% KPR - Maybe the above sentence could be rephrased to something like this?
% "The significant differences in oscillatory power across the umbra, both in temperature and velocity diagnostics, indicate rapid spatial and temporal modification of the wave propagation characteristics."  
% KPR -- I'm not sure it is proper to say that you can't reliably detect the oscillations in this location, but then say that the phases are different. Without the oscillation, the phases are undetermined. 
% For example, the oscillation in t3 at around 2100-2300 seconds seems to show a similar phase relationship to the peak oscillation region.

%The amplitude of ALMA band 3 data at quiet region sample point is also relatively weak, leaving it impossible to measure its phase relation. From the right column, we also noticed that H$\alpha$ data, though do not have a stable phase shift compare to left column, still provide some sort of oscillation signal while ALMA data shows little to no oscillation at all. Consider the formation height of ALMA band 3 and H$\alpha$, a possible assumption can be carried out that at lower chromosphere where ALMA band 3 is formed, the 3-minutes oscillation is limited in a relatively narrow and concentrated cavity, as seen on figure 6 left panel. While the wave propagating in the chromosphere towards the formation height of H$\alpha$, the cavity became wider and the component is no longer dominant by the 3-minutes signal as it is in the tight cavity that also responsible for the 3-minutes resonant oscillation.

\begin{figure*}[ht]
\centering
\includegraphics[width=0.8\linewidth]{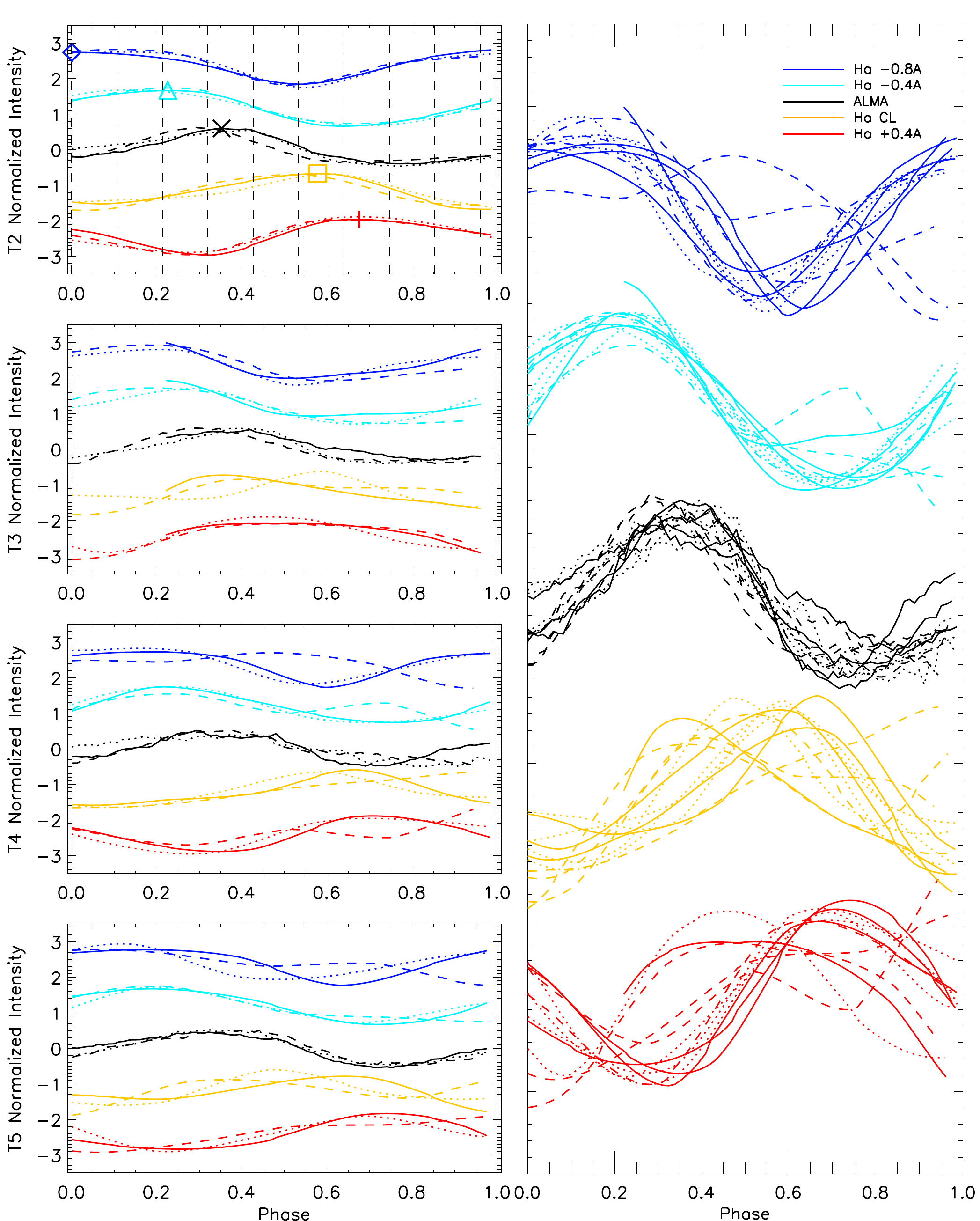}
\caption{Epoch analysis of the relative phase of the oscillation pattern in different bands using the light curves from the strong oscillation region in the left column of Figure~\ref{fig:fig7}. Left panel: analysis for each ALMA time window separately. 
%The peak positions of the $t2$ first wave were marked with different symbols to show the relatively uniform phase interval between each wavelength. 
Each curve is offset vertically for clarity.  The ALMA band 3 light curves (black) were moved to third place to best reflect the phase order. Each wave in the light curves is distinguished using different line styles (solid, dashed, and dotted line for the first, second, and third complete wave respectively). Right panel: an overlay of all waves from the left column.}
\label{fig:fig8}
\end{figure*}

In order to study the general phase relationships among all bands in the strong 3-min oscillation region, we performed an epoch analysis where individual oscillations are overlaid in terms of their phase relative to the peak of the H$\alpha$ far blue wing ($-${0.8 {\AA}}).  This was done for individual time windows $t2$ to $t5$ in the left column of Figure~\ref{fig:fig8}, with solid, dashed, and dotted lines for the first, second, and third oscillations in each time window, respectively, and for all time windows in the right column. The oscillations in different wavelengths were offset vertically for clarity, and the ALMA light curve was moved to the middle position to reflect its ordering of peak time midway between H$\alpha$ near blue wing ($-${0.4 {\AA}}, cyan) and H$\alpha$ center line (yellow).
%left column into several part using the period of wave as a time range frame and overlaid them on top of each other as shown in Figure 8. In Figure 8, we rearrange the sequence of the data so it reflect the phase shift we learned from Figure 7. For each data set in one 10-minutes scan, we only took three whole waves from one light curve and plotted them separately. The light curves of each whole wave are distinguished using line style, with solid, dotted and dashed lines represent first, second and third wave in the original light curves. We also changed the x-axis from time to phase since it can directly reveal the phase difference in percentage of one whole phase.

The left column of Figure~\ref{fig:fig8} shows that the oscillations in $t2$ are highly coherent for both the ALMA and {GST} data, reflecting the stronger oscillation power during that time window. 
%compared with the rest scans, which is coherent as we already knew that the peak 3-minutes oscillation happened at $t2$ from Figure 6. 
The peak of the first wave (solid line) in each band in $t2$ is marked with a symbol to show the progression in phase of the peaks similar to the vertical dashed lines in Figure~\ref{fig:fig7}. 
% The symbols are kind of hard to see on the plot. Is there any way they could be made bigger/thicker (e.g. SYMSIZE=2, THICK=4 if this is IDL).
% done changing the sym size/thickness
The right panel of Figure~\ref{fig:fig8}, overlaying the curves from all the time blocks, shows that this general pattern persists for all of the oscillations although there is higher variability in some bands.  The H$\alpha$ near blue wing and ALMA oscillations are particularly coherent for most of the oscillations, and there is a marked tendency for the drop in ALMA brightness (corresponding to temperature) to be steeper than its rise.
%We can also tell that the ALMA band 3 data has a more constant wave pattern than other H$\alpha$ sub bands data, as seen from the right column of Figure 8 which over-plotted all the light curves in left column. This constant behaviour can also prove that in the formation height of ALMA band 3 wavelength, the cavity is firmer compared to what in the H$\alpha$ formation height.

\section{Discussion}

In order to put the oscillations in temperature detected by ALMA into context with the wavelength-dependent brightness variations in the H$\alpha$ line, and thus account for the phase shifts in Figures~\ref{fig:fig7} and \ref{fig:fig8}, a quantitative solar atmospheric model that exemplifies the sunspot structure is needed. Inspired by \cite{2019ApJ...881...99M}, we used the RH code \citep{1991A&A...245..171R,1992A&A...262..209R,2000ApJ...536..481U} to synthesize the chromospheric radiation in 1D, which uses the Solar Irradiance Physical Modeling (SRPM) as an input for the calculation \citep{2011JGRD..11620108F}. Due to the estimated formation height of ALMA band 3 \citep{2015A&A...575A..15L}, we used a sunspot umbral model (model S) that focuses on the photosphere and chromosphere (model index 1006). The RH code solves combined equations of statistical equilibrium and radiation transfer for multi-level atoms and molecules under certain input parameters. The simulation was carried out with non-LTE setup for 6-levels of hydrogen atom (levels 0 to 5, plus the continuum), while other atomic species were treated with the LTE assumption. In addition to the default H$\alpha$ output wavelength grid, we also used the RH code to produce the radiative emission in the wavelength range from 2.6 mm to 3.4 mm corresponding to the ALMA band 3 observations.

\begin{figure*}[ht]
\centering
\includegraphics[width=0.8\linewidth]{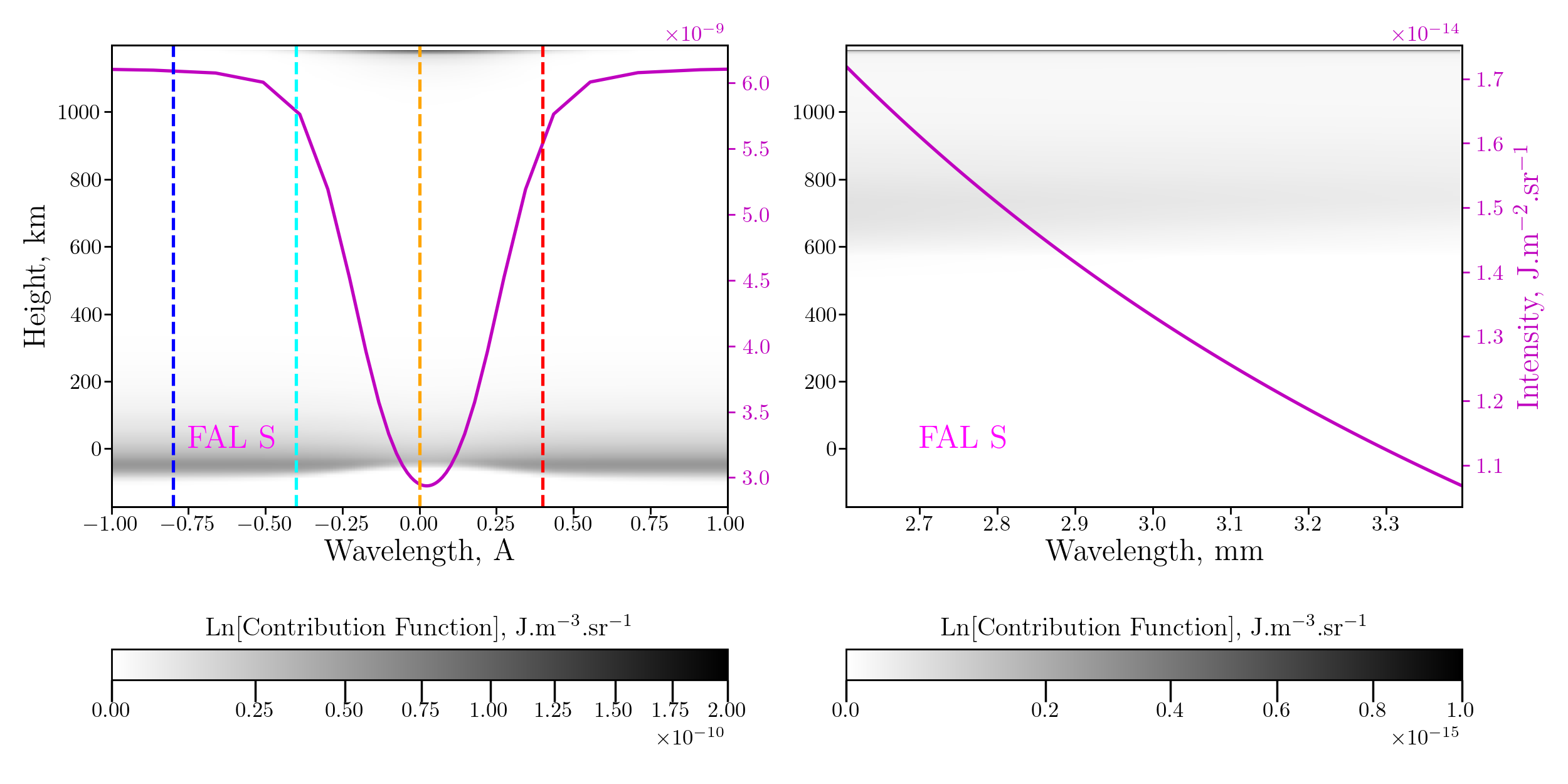}
\caption{Spectral synthesis results obtained with the RH code. Left panel: Intensity contribution function (gray shade) vs. height for the H$\alpha$ line from the FAL S model that simulates a sunspot umbra, overlaid with the emergent line profile (violet line, scale on the right axis). The wavelengths are shown relative to H$\alpha$ line center, with four vertical dashed lines to indicate the H$\alpha$ sub-bands used for the study. The line colors follow the same scheme as in Figures~\ref{fig:fig7} and \ref{fig:fig8}. Right panel: Contribution function for the emergent intensity for ALMA band 3 wavelength range from the same FAL S model, overlaid with the emergent intensity profile (violet line).}
\label{fig:fig9}
\end{figure*}

The results from the RH simulation of the umbra model are shown in Figure~\ref{fig:fig9} for both wavelength ranges. The contribution functions are illustrated in the underlying gray scale map while the emergent intensities are plotted in violet lines. The H$\alpha$ center line {position} was set to zero, and four dashed vertical lines indicate the sub-bands used in our study.
% KPR - Note that the H-alpha line profile calculated from this sunspot model using RH is much narrower than the observed H-alpha profile. The FWHM in the synthesized profile is ~0.05 nm, while from some FISS spectra, even in a sunspot the line width is about 0.1 nm. This puts the +/- 0.04 nm passbands at a ~90% "continuum" level for the synthesized profile (a location with a relatively weak slope), while in the observed profile they should fall even below the 50% level (since they are inside the FWHM), where the slope is steeper.
% KPR - I don't think this paper is the place to address sunspot models and H-alpha synthesis, but maybe this should at least be noted in the text?
From this result, we can see that the contribution function for H$\alpha$ extends over a wide height range as seen on the left panel. The major part of the center line intensity comes from a narrow layer in the middle chromosphere ($\sim$ 1100 km) while the continuum is mainly formed in the photosphere. In the ALMA band 3 wavelength range, the formation height seen in the right panel is extended between 600 km and 1000 km in the chromosphere. Combined with the phase pattern indicated by the dashed vertical lines in Figure~\ref{fig:fig7}, we can conclude that the H$\alpha$ line wing signatures are most likely dominated by the velocity of the waves, as observed by \cite{1975SoPh...41...71P}.

\begin{figure}[ht]
\centering
\includegraphics[width=0.8\linewidth]{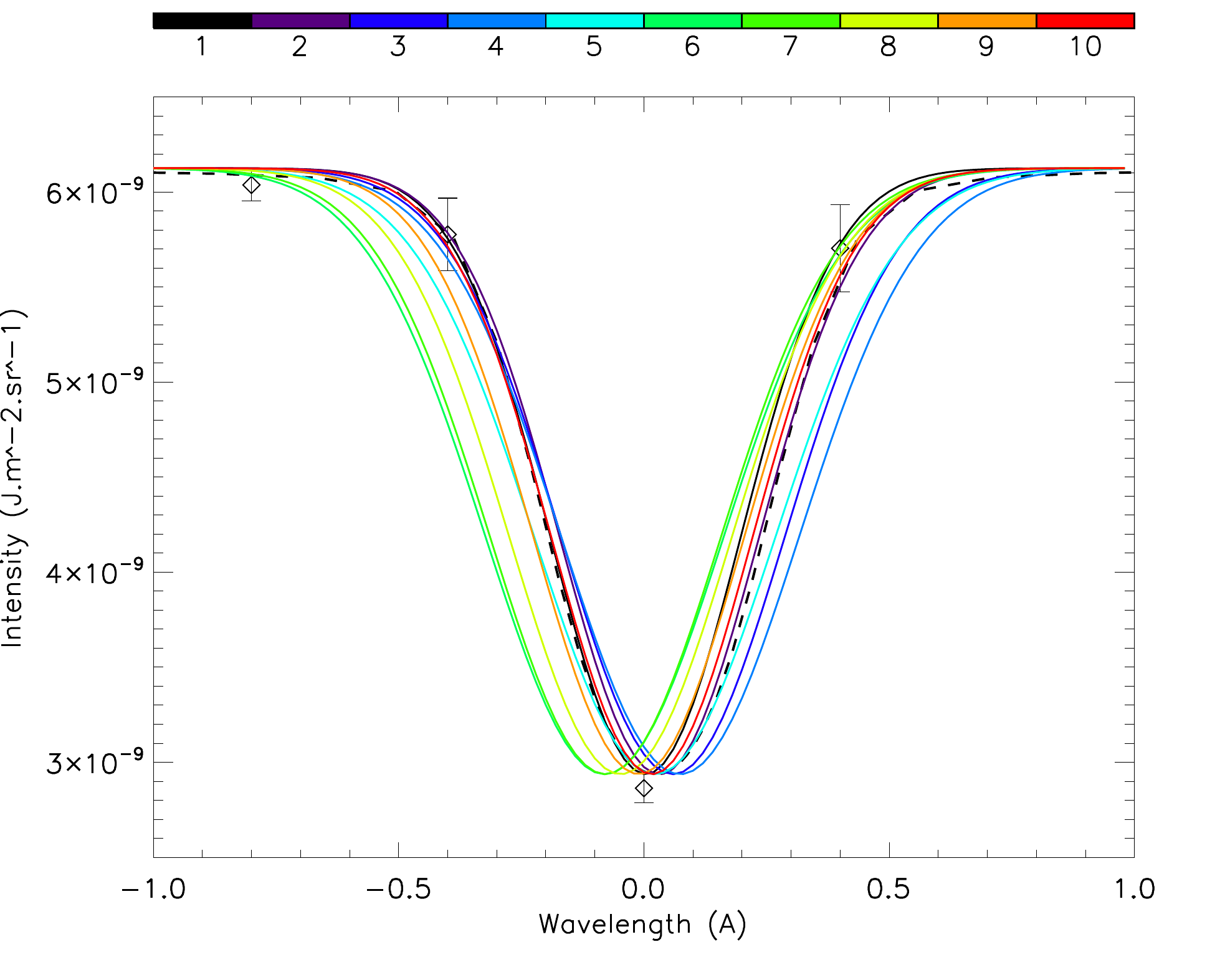}
\caption{A series of interpolations of the H$\alpha$ line profile in phase order based on the simulation result from RH code as well as the relative intensities of the four H$\alpha$ sub-bands. The dashed line represents the original emergent line profile (violet line in Figure~\ref{fig:fig9} left panel), while each colored line represents Gaussian fitting to the four H$\alpha$ line intensities at 10 times (marked with 10 vertical dashed lines in Figure~\ref{fig:fig8}). The color bar indicates the time bin. For these fits we kept the continuum and line-core intensities fixed and fit only line shift and line width. The error bars plotted for the first Gaussian fit were calculated based on the standard deviation of the light curves in the nearby quiet region for each of the four H$\alpha$ sub-bands.}
\label{fig:fig10}
\end{figure}

To better understand the influence of the oscillation on the H$\alpha$ line profile, we use the first complete oscillation in the $t2$ light curves to fit dynamic H$\alpha$ line profiles. Because no photometric calibration is available for the GST data, we instead use the normalized values from each H$\alpha$ sub-band light curve in the top-left panel of Figure~\ref{fig:fig8} to calculate the relative variations at these four sub-bands at any given time. In this process, we assume that the line profile at H$\alpha$ off-center bands can not exceed the continuum, so the light curves are obtained relative to the maximum values at these wavelengths. For the line center the relative variation is based on the mean value of its light curve. To anchor the continuum for the fits, two far off-center wavelengths (+/- {3 {\AA}}) were added to represent the continuum value, we then applied Gaussian fitting to these six points to form an interpolated line profile for a given time.  To further stabilize the fits, only line width and line center position were allowed to change during the Gaussian fitting.
% Maybe also mention that this is a slightly more robust implementation of the Dopplergram/line width measurement that people have done subtracting/adding the two H-alpha wing measurements. Actually, I'm not sure anyone has looked at the line width in the umbral oscillations, so that could be something new to point out.

The result is shown in Figure~\ref{fig:fig10}. Each color in the color bar represents one of the vertical dashed lines in the top left panel of Figure~\ref{fig:fig8}, with numbers showing the order in time. The original simulation result from the RH code is also plotted with a dashed line in the figure, and is used to provide the intensity scale shown on the vertical axis. From the color-coded time bins in the figure we can form a schematic picture of how the H$\alpha$ line profile changed with time/oscillation phase---it starts near the original dashed line profile, shifts to longer wavelengths (blue curves), then rapidly shifts to shorter wavelengths (green), and finally moves back to the center (red). 

\begin{figure}[ht]
\centering
\includegraphics[width=0.9\linewidth]{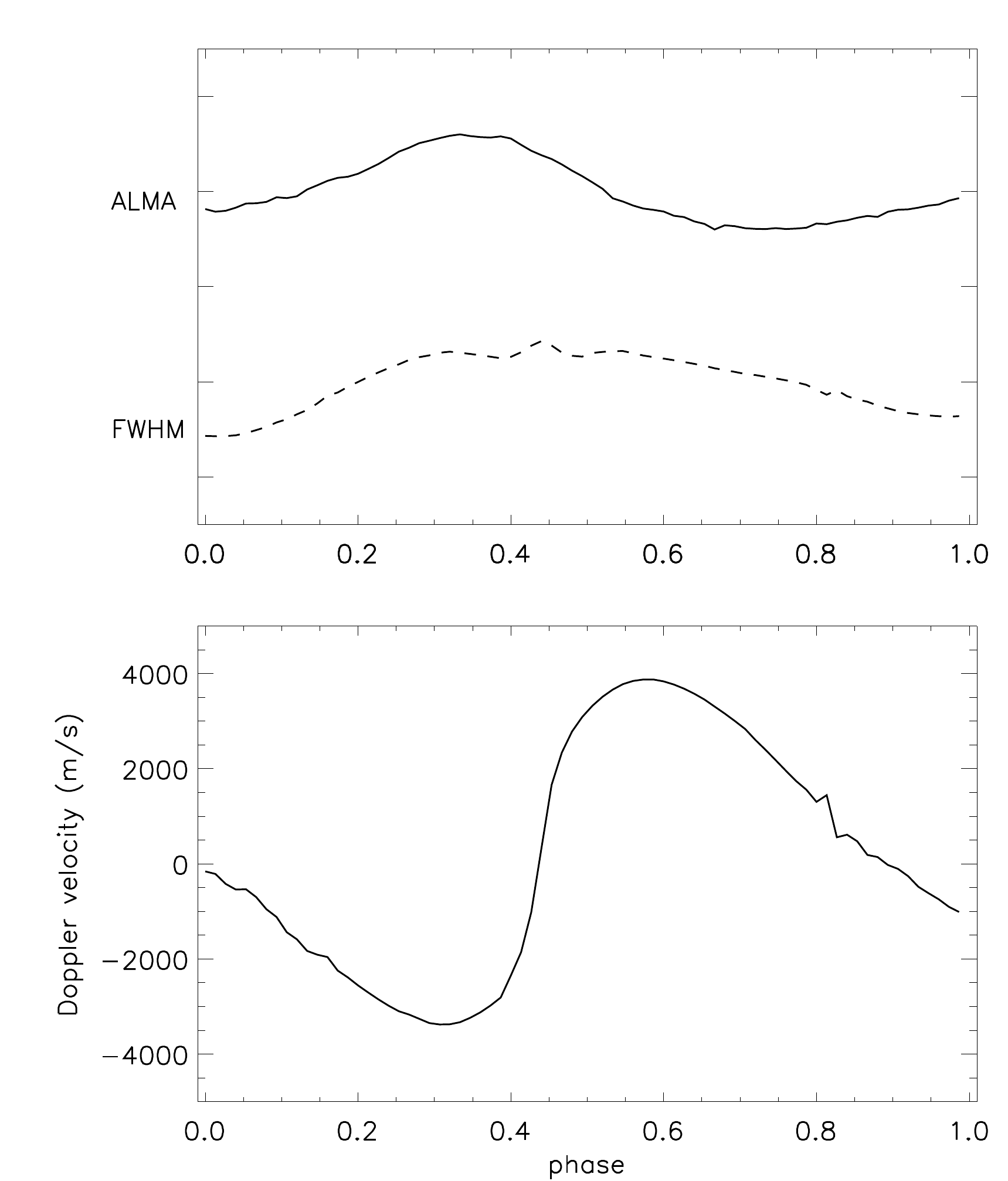}
\caption{Top panel: A comparison between an ALMA light curve (first oscillation of $t2$) and full width at half maximum (FWHM) of the Gaussian fitting in Figure~\ref{fig:fig10}. Bottom panel: The Doppler velocity corresponding to the H$\alpha$ center line shift in Figure~\ref{fig:fig10}. The upward Doppler velocity is treated as positive in the figure.}
% I think it might be nicer to just plot the FWHM and velocity fits at their full scale (on two separate plots, one above the other like this), without all the H-alpha bands, which have been extensively shown in previous figures. I would also overlay the ALMA lightcurve together with the FWHM fit curve, since they seem similar.
\label{fig:fig11}
\end{figure}

%this paragraph may need some modification and combine with the shockwave paragraph
Although these fits are approximate, the behavior of the resulting line full-width half-maximum (FWHM) and line shift (Doppler velocity) shown in Figure~\ref{fig:fig11} are suggestive. In the upper panel of Figure~\ref{fig:fig11}, we plot the temperature oscillation from ALMA alongside the line width information generated by the Gaussian fitting. Note that the line width varies in phase with the temperature variations measured by ALMA, with the greatest width corresponding to the highest temperature.  This is consistent with the finding by \cite{2019ApJ...881...99M} that ALMA temperature is correlated with H$\alpha$ line width. 

The lower panel of Figure~\ref{fig:fig11} shows the Doppler velocity. The positive Doppler shift corresponds to the upward propagating waves. The maximum downward Doppler velocity amplitude is 3.4 km/s while the maximum upward velocity is 3.9 km/s. 
% this seems like a few too many significant digits for the velocity determination...
The asymmetric Doppler velocity profile reveals that the plasma moves downward slowly with a steady acceleration and then suddenly changes its direction and moves upward with a much larger acceleration, creating a steep rise in the velocity diagram. %This profile suggests that the downward plasma motion from a previous wave packet meets the uprising motion from a successive wave packet, resulting in a rapid energy dissipation and a sudden upward acceleration and the observed transition to a strong blueshift.
% maybe - "suggests the downward plasma motion from a previous wave packet, meets the uprising motion from a successive wave packet, resulting in a rapid energy dissipation and a sudden upward acceleration and the observed transition to a strong blueshift."

The clear signature of a sawtooth pattern in the Doppler velocity for this particular oscillation, which is the largest amplitude one in our observations, has to be taken with some caution due to the relatively low cadence ($\sim$40~s) and small number of wavelengths (4) on which it is based. Even so, we find that both the Doppler velocity magnitude and trend in Figure~\ref{fig:fig11} agree well with those reported by \cite{2006A&A...456..689T} and \cite{2017ApJ...844..129C}. This asymmetrical Doppler velocity trend continues in observations of transition region lines \citep{2014ApJ...786..137T}, with a higher velocity of order 10~km s$^{-1}$. %, which can also be seen in the overlaid light curves of H$\alpha$ blue wing as well as ALMA data in Figure 8 right panel. 
Such asymmetries are evidence of steepening or shock behavior in the atmosphere of sunspot umbrae, which was discovered in the Doppler shift of Ca II \citep{1984ApJ...277..874L} as well as He I \citep{Lites1986}. %Later study also confirm the presence of such pattern in Si IV \citep{2014ApJ...786..137T} that has a Doppler velocity amplitude of $\sim$10 km/s. These studies revealed the characteristic of 3 min sunspot umbral oscillation and its connection with shock waves in the solar chromosphere, states that it may generated by the global p-mode oscillation upward leak in the photosphere, which then steepen and form shocks in the chromosphere \citep{2014ApJ...786..137T}. 
In any case, the asymmetry in the ALMA temperature behavior (slow rise in temperature followed by a faster fall) is clearly established by our observations with 2-s cadence, as seen in each of the individual oscillations in Figure~\ref{fig:fig7} and in the right panel of Figure~\ref{fig:fig8}. 
% KPR - I think it is notable that the temperature variations observed with ALMA don't show the same rapid rise, which one might expect during the abrupt heating at the time of the shock dissipation. This could be because a) the dissipation (and heating) timescales may be longer than the dynamical timescales; b) the rapid change in velocity may be a radiative transfer effect in the H-alpha line profile, related more to quick changes in the tau=1 layer at which the line signature is formed, causing a "jump" in the wave phase being sampled; c) the height of formation of ALMA changes during the heating episode causing it to sample different parts of the wave and smoothing out the heating signature. All interesting possibilities that would require future study. I guess this is also addressed below in your additional modeling, but in the regime of linear wave theory.

\begin{figure}[ht]
\centering
\includegraphics[width=0.9\linewidth]{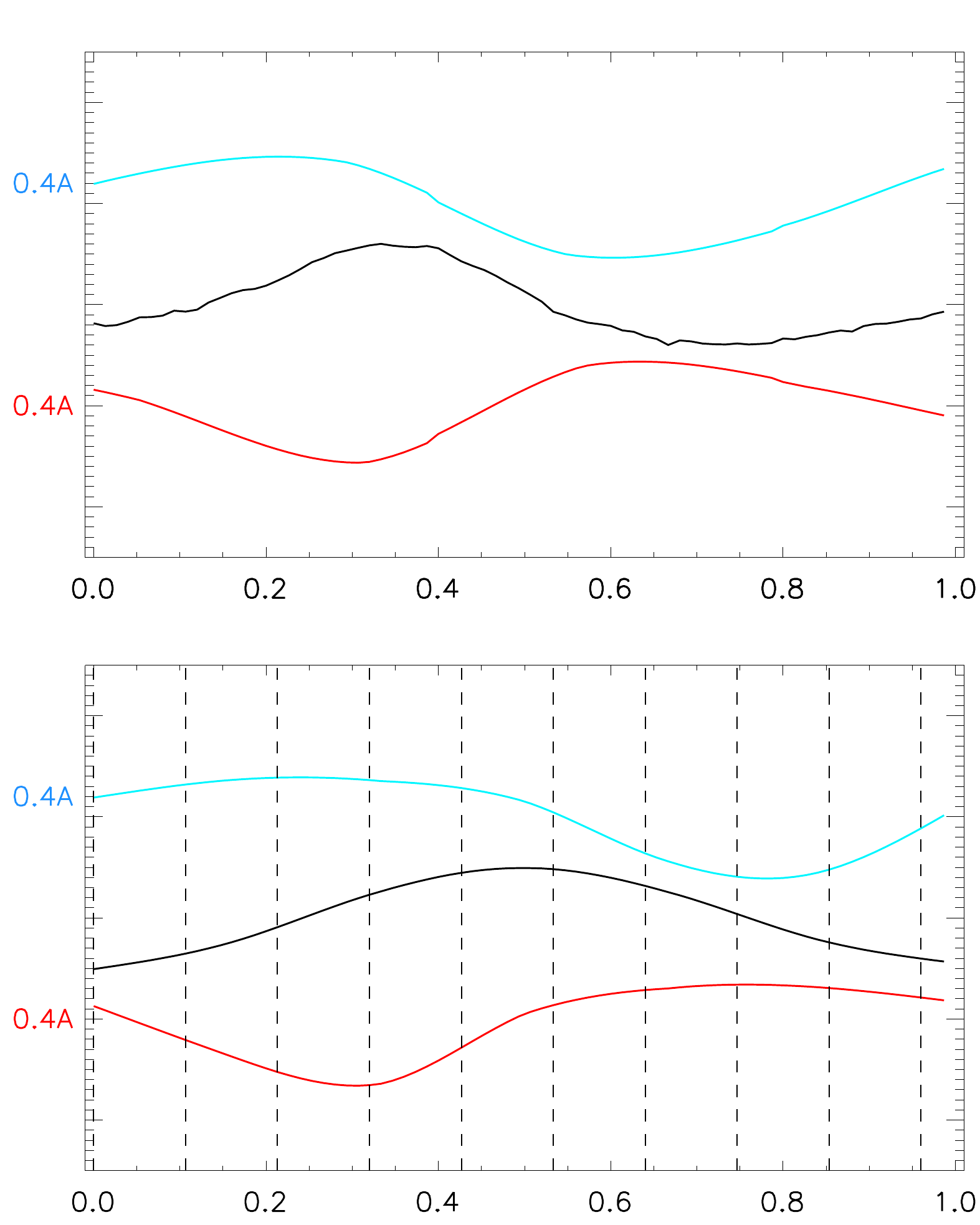}
\caption{Comparison of (a) the observed H-alpha intensity in red and blue wing and ALMA temperature variation with (b) the same quantities from the hydrodynamic acoustic model of \cite{2015ApJ...808..118C}. Vertical dashed lines mark positions in phase similar to those in the top left panel of Figure~\ref{fig:fig8}}
\label{fig:fig12}
\end{figure}

The ALMA temperature variation---its magnitude, phase relative to H$\alpha$ Doppler velocity, and its asymmetric shape---all provide new information on the phenomenon of umbral 3-min oscillations. We have investigated whether these observations can be explained by current theories of acoustic wave propagation by performing a simulation from the model of \cite{2015ApJ...808..118C}.
%, the details of which will be given in a future paper.  Here we simply
We compare the results of our simulations from their model with the observations, as shown in Figure~\ref{fig:fig12}. In brief outline, the steps we took to produce Figure~\ref{fig:fig12}b are (i) implementing their model in Python, (ii) using it to calculate the atmospheric parameter perturbations (velocity $v$, number density perturbation $dn_e$, and temperature perturbation $dT$) as a function of height and time, (iii) modifying the FAL umbral model by adding these perturbations as fractional changes at multiple time steps, and (iv) running the RH code at these multiple time steps to simulate the H$\alpha$ line profile and ALMA brightness temperature. The good agreement of the phase differences between the observations (Figure~\ref{fig:fig12}a) and simulations (Figure~\ref{fig:fig12}b) suggest that the ALMA temperature variations are mainly a consequence of an upward propagating acoustic wave in the stratified medium of the sunspot umbra. 
%The symmetric temperature variation from the model is due to the linear theory used for wave propagation in \cite{2015ApJ...808..118C}. The theory was updated for nonlinear (shock) behavior by \cite{2017ApJ...844..129C}, which we suggest can account for the saw-tooth asymmetry in the observations although we have not yet done those simulations.
% it looks like the observed ALMA temperature peaks at a phase of about 0.4, roughly coincident with the strong velocity reversal. Whereas the simulated wave signatures place the temperature peak roughly halfway between the two (symmetric) velocity maxima (as expected for simple acoustic wave propagation. 

%Unlike other wavelength, the Doppler velocity shift of H$\alpha$ only shows a magnitude of 4 km/s based on our study, which is comparable with the 5 km/s maximum given in \cite{Tziotziou2006}'s paper. However we can still calculate if it is comparable with the temperature variation observed by ALMA. During the plasma falling process indicate by the increasing doppler velocity, We noticed that there is a peak in ALMA band 3 light curve, as well as in the FWHM of the Gaussian fitting. The line broadening could represent a signature of heating in the plasma since the increasing brightness temperature is also appeared in ALMA data. If we assume free-fall and adiabatic compression of pure hydrogen case for a simple calculation, we can find that the gravitational potential would generate an increase of $\sim$ 3000 K in temperature, which is far enough considering that the brightness temperature increasing in ALMA band 3 is only $\sim$ 880 K.

%comparison between doppler velocity and typical acoustic velocity may be needed.

\begin{figure}[ht]
\centering
\includegraphics[width=0.9\linewidth]{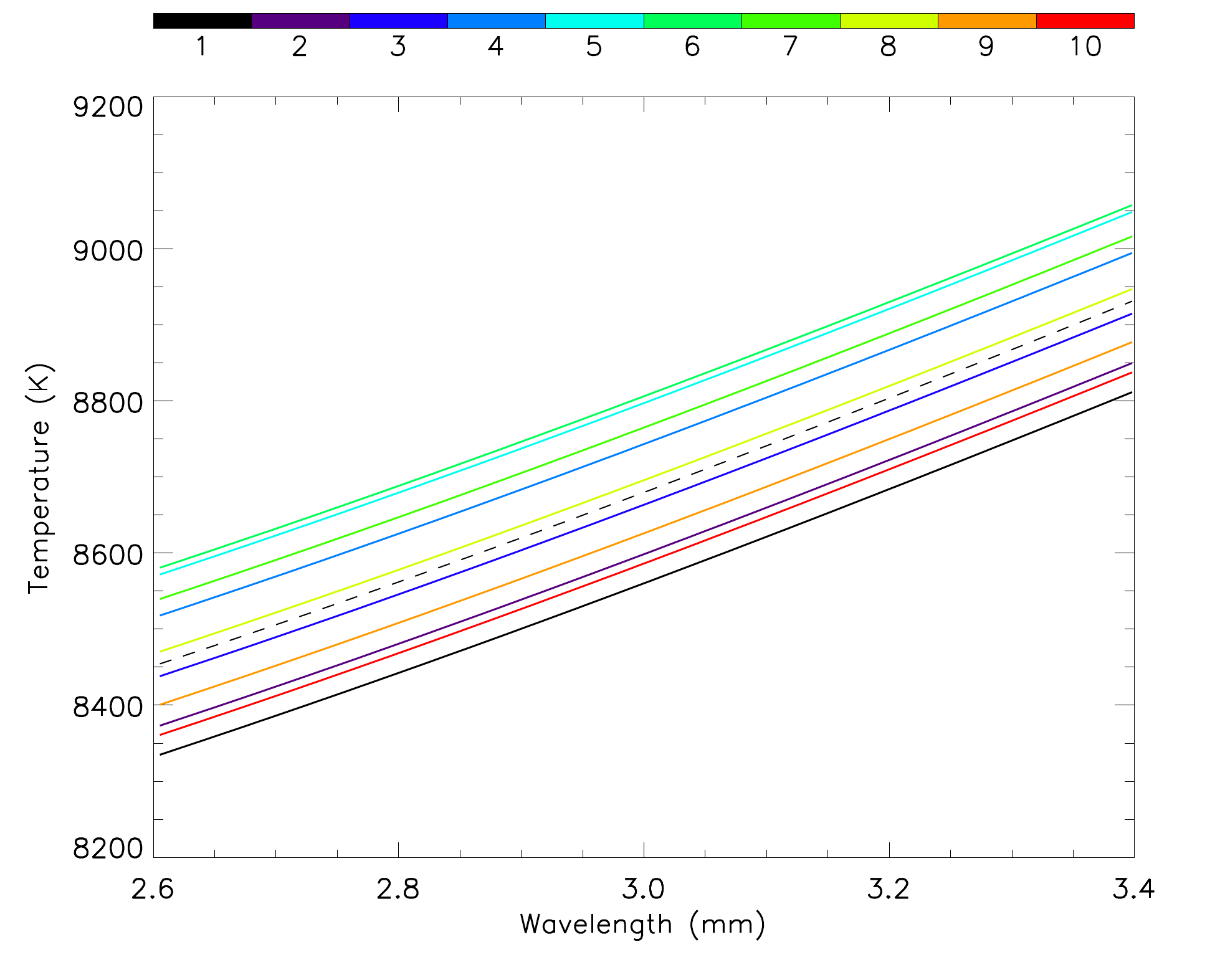}
\caption{A series of ALMA temperature in phase order based on the simulation result from RH code. The dashed line represents the original temperature profile calculated based on the emergent line profile (purple line) in Figure~\ref{fig:fig8} right panel. The color bar indicates the time bin. The temperature gradient reflects the fact that shorter wavelength of ALMA band3 sees deeper in the chromosphere.}
\label{fig:fig13}
\end{figure}

Similar to Figure~\ref{fig:fig10}, we plotted the temperature profile based on the simulated emergent intensity in band 3 wavelengths from the RH code at the 10 phases marked in Figure~\ref{fig:fig12}b. The colors from blue to red indicate increasing phase (or time), as shown in the color bar. The temperature profile starts at the lowest temperature, rises to the peak, and then falls back. The rapid line shift that happens between time bins 4 to 7 in Figure~\ref{fig:fig10} corresponds to the temperature peak in Figure~\ref{fig:fig13}, but the earlier peak in the observations (Figure~\ref{fig:fig12}a) relative to the linear model (Figure~\ref{fig:fig12}b) suggests a tendency for the chromosphere to be heated more strongly at this time than described by the linear theory. The absolute temperature and temperature fluctuation predicted by the RH code, about 8700 K and $\pm1100$ K, respectively at 3 mm, are both somewhat higher than the observations (Figure~\ref{fig:fig7}), which are around 8000 K and $\pm500$ K.  See \citet{2017ApJ...850...35L} for a detailed comparison of the brightness temperatures with various chromospheric models for this same sunspot observed one day earlier.

%(Check how the Bifrost model does, from Loukitcheva et al. 2015)

\section{Conclusions}

Our study describes the first observations of sunspot oscillations detected by ALMA. The ALMA observations provide unique measurements of the chromospheric temperature variation tied to electron temperature via free-free emission, and so is relatively insensitive to atomic abundances and local non-equilibrium conditions.

We applied Fourier analysis to the ALMA band 3 data to create spatial maps of the 3-min power amplitude.  The 3-min power maps correlate well to similar maps for all H$\alpha$ sub-bands except the far red wing ({+0.8 {\AA}}) where the 3-min oscillation signal is weak.  From the light curves in Figures 7 and 8, we discovered a relatively constant phase relationship among H$\alpha$ sub-band intensity and ALMA temperature as well as evidence for a sawtooth pattern in the ALMA temperature variation, albeit less abrupt than the velocity transition seen in H$\alpha$.
The phase relationship among the different H$\alpha$ bands sampling different parts of the line profile is consistent with a shift of the line center during the oscillation as well as a slight variation in line width. The ALMA temperature and the H$\alpha$ line width peak at around the same time, but the ALMA temperature shows a more rapid fall after the peak time.

We used RH code to simulate the H$\alpha$ line profile and the expected ALMA emission based on the FAL S sunspot atmosphere model. The results are generally in agreement with observations although the ALMA band 3 brightness temperature predicted by the FAL S model (8700 K) is some 9\% greater than observed (8000 K). In order to understand the phase relationship of the observed ALMA temperature variation, we used the linear theory of \cite{2015ApJ...808..118C} for the propagation of a hydrodynamic disturbance in a simplified solar atmosphere to obtain the velocity, density, and pressure perturbations as a function of height, then applied those perturbations to the FAL S model.  Calculations of the H$\alpha$ and ALMA emission using repeated runs of the RH code for different times during one oscillation of such a disturbance gave results that are in reasonable agreement with the model. In particular, the calculated relative phase relationships among the H$\alpha$ near blue wing intensity, ALMA band 3 temperature, and H$\alpha$ near red wing intensity are close to those observed, showing that the ALMA temperature variations are largely due to atmospheric heating expected for a propagating disturbance. 

{The symmetric temperature variation from the model is due to the linear theory used for wave propagation in \cite{2015ApJ...808..118C} which} cannot reproduce the slight sawtooth pattern observed in both the ALMA band 3 temperatures and H$\alpha$ Doppler velocity, therefore a study of the non-linear solution for the wave function is much needed. Based on previous research \citep{2017A&A...599A..15L, 2017ApJ...844..129C}, the non-linear effect can provide not only a physics scenario of the wavefront steepening and shock formation that is related with the sawtooth pattern, but also an explanation for the 2nd harmonic seen in the umbra power spectrum in Figure~\ref{fig:fig4}. {We also note that the RH code gives an H$\alpha$ line width for the FAL S model that is significantly narrower than commonly observed \citep{1965ZA.....63...35F,2005asus.book.....W}. We have done a preliminary study of other models (e.g. the FAL penumbral model R and the Maltby model \citep{1986ApJ...306..284M}, which can correct the H$\alpha$ line-width discrepancy but results in estimated ALMA temperatures (and heights of formation) that are much too high. Here we simply note the discrepancy---a more thorough investigation will be addressed in a future paper now in preparation.}

These new observations show the power of ALMA mm-wave observations for providing entirely new diagnostics of the solar atmosphere.  Because of the low solar activity since these science-verification observations were made, no additional observations of sunspot umbrae have been possible to date.  However, advances in ALMA capabilities for solar observations, such as higher spatial resolution, polarization, and additional frequency bands that provide diagnostics at both lower and higher heights, all offer great promise for investigations in the near future as solar activity returns.  Such observations should motivate more sophisticated modeling that includes the effect of the hydrodynamic waves on ionization states as well as the effect of the strong sunspot magnetic fields.

%% The reference list follows the main body and any appendices.
%% Use LaTeX's thebibliography environment to mark up your reference list.
%% Note \begin{thebibliography} is followed by an empty set of
%% curly braces.  If you forget this, LaTeX will generate the error
%% "Perhaps a missing \item?".
%%
%% thebibliography produces citations in the text using \bibitem-\cite
%% cross-referencing. Each reference is preceded by a
%% \bibitem command that defines in curly braces the KEY that corresponds
%% to the KEY in the \cite commands (see the first section above).
%% Make sure that you provide a unique KEY for every \bibitem or else the
%% paper will not LaTeX. The square brackets should contain
%% the citation text that LaTeX will insert in
%% place of the \cite commands.

%% We have used macros to produce journal name abbreviations.
%% \aastex provides a number of these for the more frequently-cited journals.
%% See the Author Guide for a list of them.

%% Note that the style of the \bibitem labels (in []) is slightly
%% different from previous examples.  The natbib system solves a host
%% of citation expression problems, but it is necessary to clearly
%% delimit the year from the author name used in the citation.
%% See the natbib documentation for more details and options.

\begin{acknowledgments}

{This paper makes use of the following ALMA data: ADS/JAO.ALMA\#2011.0.000020.SV. ALMA is a partnership of ESO (representing its member states), NSF (USA) and NINS (Japan), together with NRC (Canada), MOST and ASIAA (Taiwan), and KASI (Republic of Korea), in cooperation with the Republic of Chile. The Joint ALMA Observatory is operated by ESO, AUI/NRAO and NAOJ.}

{BBSO operation is supported by NJIT and US NSF AGS-1821294 grant. GST operation is partly supported by the Korea Astronomy and Space Science Institute, the Seoul National University, and the Key Laboratory of Solar Activities of Chinese Academy of Sciences (CAS) and the Operation, Maintenance and Upgrading Fund of CAS for Astronomical Telescopes and Facility Instruments.}

{The National Radio Astronomy Observatory is a facility of the National Science Foundation operated under cooperative agreement by Associated Universities, Inc.  YC gratefully acknowledges support from the NRAO Student Observing Support program.}

{The data used in this paper can be obtained from the following sources: The IRIS data can be obtained from https://iris.lmsal.com/data.html.  GST data can be obtained from the BBSO data request form http://www.bbso.njit.edu/$\sim$vayur/nst\_requests2/.  SDO data (AIA and HMI) can be obtained from http://jsoc.stanford.edu/.}

\end{acknowledgments}

%% This command is needed to show the entire author+affilation list when
%% the collaboration and author truncation commands are used.  It has to
%% go at the end of the manuscript.
%\allauthors

%% Include this line if you are using the \added, \replaced, \deleted
%% commands to see a summary list of all changes at the end of the article.
%\listofchanges

\end{document}